\documentclass{emulateapj} 

\received{April 11,2014}
\accepted{June 2, 2014}

\usepackage{hyperref}
\input{colordvi}
\hypersetup{
    bookmarks=true,         
    unicode=false,          
    pdftoolbar=true,        
    pdfmenubar=true,        
    pdffitwindow=true,     
    pdfstartview={FitH},    
    pdftitle={BD+44 HST spectra},    
    pdfauthor={vmplacco},     
    pdfsubject={Astronomy},   
    pdfcreator={dvipdf},   
    pdfproducer={dvipdf}, 
    pdfkeywords={metal-poor stars},
    pdfnewwindow=true,      
    colorlinks=true,       
    linkcolor=red,          
    citecolor=blue,        
    filecolor=magenta,      
    urlcolor=cyan,           
    breaklinks=true,
    linktocpage
}

\newcommand{\bd}{\object{BD+44$^\circ$493}}
\newcommand{\eps}[1]{\mbox{log~$\epsilon$\,(#1)}}
\newcommand{\metal}{[Fe/{}H]}
\newcommand{\cfe}{[C/{}Fe]}
\newcommand{\abund}[2]{[#1/{}#2]}

\newcommand{\teff}{$T_{\rm eff}$\,}
\newcommand{\logg}{log\,$g$\,}

\shorttitle{\bd\ HST Spectra}
\shortauthors{Placco et al.}

\begin{document}

\title{\textit{Hubble Space Telescope} Near-Ultraviolet Spectroscopy of the \\
Bright CEMP-no Star \bd\footnotemark[1]}

\footnotetext[1]{
Based on observations made with the NASA/ESA Hubble Space Telescope, 
obtained at the Space Telescope Science Institute, 
which is operated by the Association of Universities for Research in 
Astronomy, Inc., under NASA contract NAS~5-26555. 
These observations are associated with program GO-12554, and we also make use of
data taken in program GO-12268.}

\author{
Vinicius  M. Placco\altaffilmark{2},
Timothy    C. Beers\altaffilmark{3,4},
Ian     U. Roederer\altaffilmark{5},
John       J. Cowan\altaffilmark{6},
Anna         Frebel\altaffilmark{7},\\
Dan          Filler\altaffilmark{8},
Inese      I. Ivans\altaffilmark{8},
James     E. Lawler\altaffilmark{9},
Hendrik      Schatz\altaffilmark{4,10},
Christopher  Sneden\altaffilmark{11},\\
Jennifer  S. Sobeck\altaffilmark{12},
Wako           Aoki\altaffilmark{13},
Verne      V. Smith\altaffilmark{3}
}

\altaffiltext{2}{Gemini Observatory,
                 Hilo, HI 96720, USA}
\altaffiltext{3}{National Optical Astronomy Observatory, 
                 Tucson, AZ 85719, USA}
\altaffiltext{4}{JINA: Joint Institute for Nuclear Astrophysics}
\altaffiltext{5}{Department of Astronomy, University of Michigan,
                 Ann Arbor, MI 48109, USA}
\altaffiltext{6}{Homer L. Dodge Department of Physics and Astronomy, University
                 of Oklahoma, Norman, OK 73019, USA}
\altaffiltext{7}{Kavli Institute for Astrophysics and Space Research and
                 Department of Physics, Massachusetts Institute of Technology, 
				 Cambridge, MA 02139, USA}
\altaffiltext{8}{Department of Physics and Astronomy, The University of Utah,
                 Salt Lake City, UT 84112, USA}
\altaffiltext{9}{Department of Physics, University of Wisconsin, 
                 Madison, WI 53706, USA}                                                
\altaffiltext{10}{National Superconducting Cyclotron Laboratory, Michigan State
                 University, East Lansing, MI 48824, USA} 
\altaffiltext{11}{Department of Astronomy and McDonald Observatory, University
                  of Texas, Austin, TX 78712, USA}
\altaffiltext{12}{Department of Astronomy, University of Virginia,
                  Charlottesville, VA 22904, USA}
\altaffiltext{13}{National Astronomical Observatory of Japan, 2-21-1 Osawa,
                  Mitaka, Tokyo 181-8588, Japan}

\addtocounter{footnote}{13}

\begin{abstract}

We present an elemental-abundance analysis, in the near-ultraviolet (NUV)
spectral range, for the extremely metal-poor star \bd, a 9th magnitude subgiant
with [Fe/H] $= -3.8$ and enhanced carbon, based on data acquired with the Space
Telescope Imaging Spectrograph on the \textit{Hubble Space Telescope}. This star
is the brightest example of a class of objects that, unlike the great majority
of carbon-enhanced metal-poor (CEMP) stars, does not exhibit over-abundances of
heavy neutron-capture elements (CEMP-no).  In this paper, we validate the
abundance determinations for a number of species that were previously studied in
the optical region, and obtain strong upper limits for beryllium and boron, as
well as for neutron-capture elements from zirconium to platinum, many of which
are not accessible from ground-based spectra. The boron upper limit we obtain
for \bd, \eps{B}$ < -0.70$, the first such measurement for a CEMP star, is the
lowest yet found for very and extremely metal-poor stars. In addition, we
obtain even lower upper limits on the abundances of beryllium, \eps{Be}$ <
-2.3$, and lead, \eps{Pb} $< -0.23$ ([Pb/Fe]$ < +1.90$), than those reported
by previous analyses in the optical range. Taken together with the
previously measured low abundance of lithium, the very low upper limits on
Be and B suggest that \bd\ was formed at a very early time, and that it could well
be a bona-fide second-generation star. Finally, the Pb upper limit
strengthens the argument for non-$s$-process production of the heavy-element
abundance patterns in CEMP-no stars.  

\end{abstract}

\keywords{Galaxy: halo---techniques: spectroscopy---stars: 
abundances---stars: atmospheres---stars: Population II---stars:
individual (\bd)}

\section{Introduction}
\label{intro}

Carbon-enhanced metal-poor (CEMP) stars are a subset of metal-poor 
\citep[MP; \metal\footnote{\abund{A}{B} =
$\log(N_A/{}N_B)_{\star} - \log(N_A/{}N_B) _{\odot}$, where $N$ is the
number density of atoms of a given element, and the indices refer to the
star ($\star$) and the Sun ($\odot$).}$ <-$1.0, e.g.,][]{beers2005,
frebel2011} and very metal-poor (VMP; [Fe/H] $ < -2.0$) stars that
exhibit elevated carbon relative to iron, [C/Fe], sometimes referred to
as carbonicity \citep[\cfe $\geq +$1.0;][]{placco2011}. It has recently
been recognized that a more appropriate division on [C/Fe] for the
identification of CEMP stars is at somewhat lower carbonicity, e.g.,
[C/Fe] $ \geq +$0.7 \citep{aoki2007,carollo2012,norris2013}. In the past
few decades, it has become clear that such stars comprise a significant
fraction of all VMP stars ($\sim$ 10-20\%; \citealt{beers1992,
norris1997,rossi1999,beers2005,cohen2005, marsteller2005,rossi2005,
frebel2006,lucatello2006,norris2007, carollo2012,cohen2013,norris2013,
spite2013}), one that increases strongly with declining metallicity,
from 30\% for [Fe/H] $< -$3.0, to 40\% for [Fe/H] $< -$3.5, $\sim
$75\% for [Fe/H] $< -$4.0 and 100$\%$ for [Fe/H]$<-$5.0. 
This trend has been confirmed with the many
thousands of CEMP stars identified by \citet{lee2013} from the Sloan
Digital Sky Survey \citep[SDSS;][]{york2000}, and its extensions SEGUE-1
\citep[Sloan Extension for Galactic Exploration and Understanding;][]{yanny2009}
and SEGUE-2 (C. Rockosi et al., in preparation).

For most CEMP stars there exists a clear correlation between carbon
enhancement and the over-abundance of elements produced by $s$-process
nucleosynthesis, such as Ba \citep[CEMP-$s$ stars; see][]{beers2005}. This
behavior is consistent with the hypothesis that these enhancements (both
for carbon and elements produced by the $s$-process) are due to
nucleosynthesis processes that took place during the asymptotic
giant-branch (AGB) stage of stellar evolution \citep[e.g.,][]{herwig2005,
sneden2008}. The resulting abundance pattern can arise from
the star itself \citep[which should rarely be
found, but see][]{masseron2006} or by a now-extinct binary companion
that transferred material to the surviving (observed) companion
\citep{stancliffe2008}. Multi-epoch radial-velocity measurements by 
\citet{mcclure1983}, \citet{mcclure1990}, and \citet{jorissen1998} 
demonstrated that the frequency of detected binaries among equivalent 
Population~I Ba stars and Population~II CH stars indicated that
essentially all are members of binary systems. \citet{lucatello2005}
conducted a similar study for members of the more metal-deficient
CEMP-$s$ sub-class of stars, and reached the same conclusion.

An intriguing variation on this behavior was initially recognized by
\citet{barbuy1997} and \citet{hill2000}. The CEMP stars CS~22948-027 and
CS~29497-034 were found not only to be rich in the elements commonly
produced by $s$-process nucleosynthesis, such as Sr, Y, Ba, and La, but
also in Eu, an element that, for extremely low-metallicity stars, is
more likely produced by $r$-process nucleosynthesis. Additional studies
by many groups have now identified $\sim$50 of these so-called CEMP-r/s
stars. These cases, once thought to be the rare exceptions, are as
commonly represented among CEMP stars as the ``$s$-only'' variety. The
origin of the abundance patterns of the CEMP-$r/s$ stars is not yet
clear, and many scenarios have been proposed \citep{jonsell2006,
masseron2010,lugaro2012}. For example, an association with a $^{22}$Ne
neutron source in intermediate-mass AGB stars has been suggested for
progenitors of CEMP-$r/s$ stars, rather than the $^{13}$C neutron source
thought to be active for low-mass AGB stars, the likely progenitors of
the CEMP-s stars (\citealt{placco2013}, Hollek et al. 2014, submitted). Mass 
transfer from a companion that passed through the AGB phase of stellar
evolution has been suggested previously as an explanation for the CEMP-$s$ 
and CEMP-$r/s$ classes of stars, based on the high fraction of such stars 
found in binary systems \citep{masseron2010,allen2012,bisterzo2012}.

The story has become richer still. \citet{aoki2007}, and others since,
have shown that the correlation between carbon enhancement and the
over-abundances of $s$- or $r/s$-elements no longer persists (or at
least is different in nature) for the majority of CEMP stars with [Fe/H]
$<-$2.7. These so-called CEMP-no stars (indicating no enhancement of
neutron-capture elements) suggest that a variety of carbon-producing
mechanisms, other than that associated with AGB stars, may have played a
role in the early universe. Possible progenitors for this sub-class
include massive, rapidly rotating, mega metal-poor (MMP; [Fe/H] $<-$6.0)
stars, sometimes referred to as ``spinstars'' \citep{chiappini2013},
which models suggest have greatly enhanced abundances of CNO due to
distinctive internal burning and mixing episodes, followed by strong
mass loss \citep{meynet2006, meynet2010,hirschi2007}. Another possible
scenario is pollution of the ISM by so-called faint supernovae
associated with the first generations of stars, which experience
extensive mixing and fallback during their explosions \citep{umeda2005,
tominaga2007}. Although more data are desired for CEMP-no stars,
\citet{hansen2013} report that the fraction of binaries among stars
within this sub-class is no higher than expected for random samples of
VMP giants. \citet{cohen2013}, \citet{norris2013},
\citet{starkenburg2014}, and J.\ Andersen et al.\ (in preparation) reach
similar conclusions. Thus, contribution of material from an evolved
binary companion is apparently not required in order to form CEMP-no stars.

The recently reported extremely metal-poor Damped Lyman-$\alpha$ system
by \citet{cooke2011} ([Fe/H]$\sim-$3.0) exhibits enhanced carbon ([C/Fe]
= $+$1.5) and other elemental abundance signatures that
\citet{kobayashi2011} also associate with production by faint
supernovae. This observation is suggestive of similar carbon-production
and enrichment mechanisms in the early universe -- both locally and in
high-redshift systems. It is presumably no coincidence \citep{beers2005,
frebel2007b} that five of the six stars known with [Fe/H] $ < -4.5$ are
confirmed CEMP-no stars \citep{christlieb2002,frebel2005,norris2007,
caffau2011a,hansen2014,keller2014}.

There are two observational keys required to advance our understanding of these
ancient stars and how they are related to early Galactic chemical evolution. One
is to obtain the full set of C, N, and O abundances for as many CEMP stars as
possible, an activity that is being pursued by a number of groups (see, e.g.,
\citealt{kennedy2011,placco2013,placco2014, roederer2014}; C.\ Hansen et al., in
preparation; C.\ Kennedy et al., in preparation). The other is to obtain as
complete an inventory as possible of the light and heavy neutron-capture
elements for representative examples of the known varieties of CEMP stars.
Although we have partial information from previous ground-based high-resolution
spectroscopic observations, there remain many key elements, such as the light
species Ge and Zr, and heavier species such as Os and Pt, that can only be
obtained through near-ultraviolet (NUV) spectroscopy
\citep[e.g.,][]{sneden1998,cowan2005}. The element Pb is of particular
importance, as it may provide a useful discriminant between a number of possible
nucleosynthesis pathways \citep{busso1999,cohen2006, ito2013}.  

Since it was installed on board the \textit{Hubble Space Telescope}
(\textit{HST}) in 1997, the Space Telescope Imaging Spectrograph (STIS)
has been the only instrument available for the high-resolution NUV
spectroscopy required to make these measurements. We have recently
completed a new \textit{HST}/STIS observing program to collect
high-quality NUV spectroscopy for three CEMP stars, including one member
of each of the CEMP-no, CEMP-$s$, and CEMP-$r/s$ sub-classes. In this
paper, we perform an abundance analysis of high-resolution NUV
spectroscopy for the star \bd, the brightest known member of the
sub-class of CEMP-no stars. We fill in the abundance patterns, as best
as possible, for elements beyond the iron peak, including eight species
not accessible from ground-based observations. Section \ref{obssec}
describes our observations and reductions, and compares abundances
derived from NUV lines with those derived previously from optical lines
by \cite{ito2013}. Section \ref{absec} describes our abundance analysis
in detail. We present a brief discussion and our conclusions in Section
\ref{diss}.

\vspace{0.5cm}

\begin{figure*}[!ht]
\epsscale{1.09}
\plotone{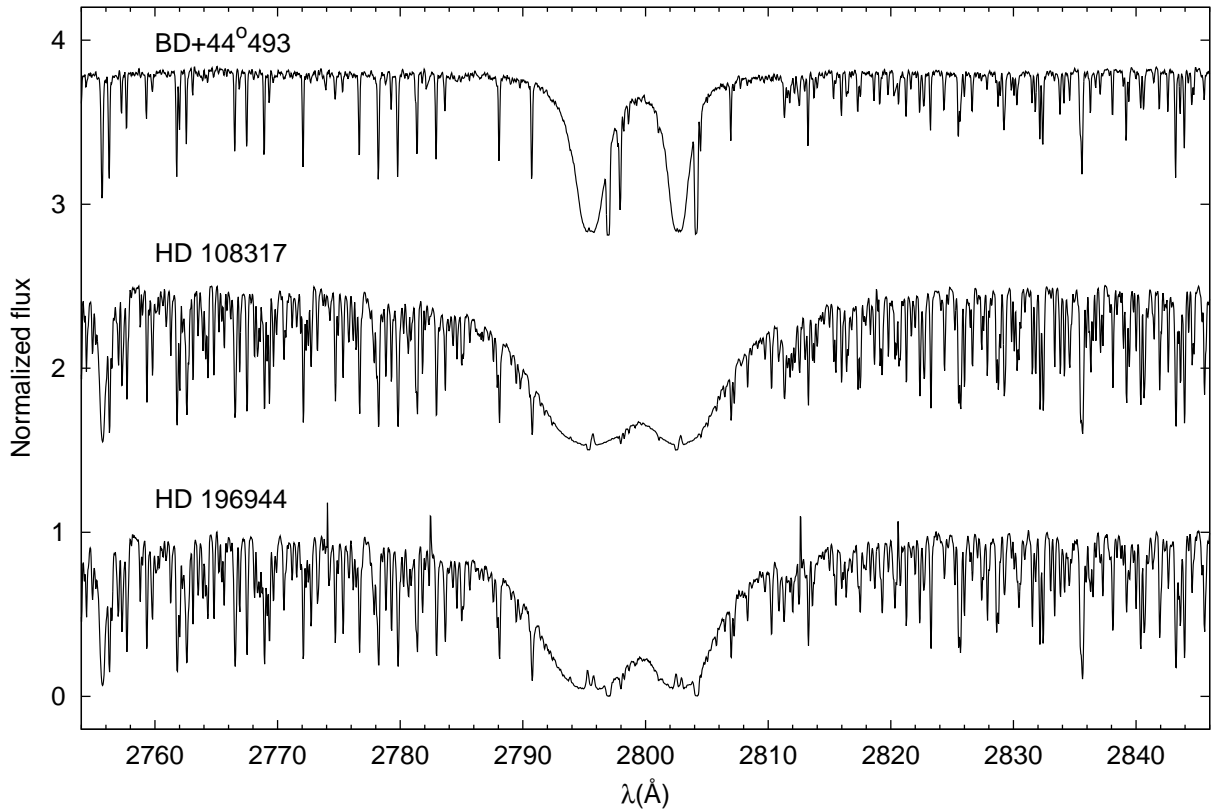}
\caption{\textit{HST}/STIS spectra for BD+44$^\circ$493, 
HD~108317, and HD~196944, in the region of the
\ion{Mg}{2} doublet at 2800\,\AA.  All three stars have similar
\teff and \logg; the substantially lower metallicity of BD+44$^\circ$493 is 
immediately apparent.}
\label{mg_comp}
\end{figure*}

\section{Observations and Measurements}
\label{obssec}

\subsection{\textit{HST}/STIS Spectra}

New STIS \citep{kimble1998,woodgate1998} observations of \bd\ were
obtained as part of Program GO-12554, using the E230M echelle grating,
centered on 2707\,{\AA}, and the NUV Multianode Microchannel Array
detector. There were two observational sequences of four
individual exposures, taken on February 28, 2012. Each sequence had an
exposure time of 2 hours and 52 minutes, with a total integration 
time of 5 hours and 44 minutes. The
0\farcs06\,$\times$\,0\farcs2 slit yields a $\sim$~2-pixel resolving
power (R~$\equiv \lambda/\Delta\lambda$) $\sim$~30,000. Our setup
produced a wavelength coverage from 2280\,{\AA}--3070\,{\AA} in a single
exposure. The observations were reduced and calibrated using the
standard \textit{calstis} pipeline. The S/N of the combined spectrum
varies from $\sim$50~pix$^{-1}$ near 2300\,{\AA}, to $\sim$80~pix$^{-1}$
near 2700\, {\AA}, to $>$100~pix$^{-1}$ near 3070\, {\AA}.

\begin{figure}[!ht]
\epsscale{1.15}
\plotone{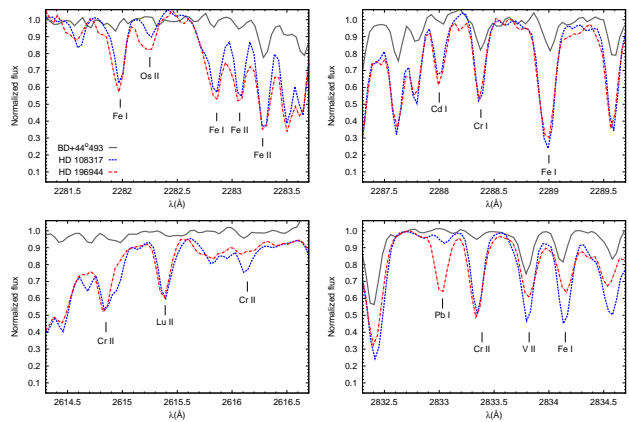}
\caption{\textit{HST}/STIS spectra for BD+44$^\circ$493, 
HD~108317, and HD~196944,
in the regions of the lines of \ion{Os}{2},
\ion{Cd}{1}, \ion{Lu}{2}, and \ion{Pb}{1}. The contrast in the
abundances of these neutron-capture elements for the latter two stars,
relative to BD+44$^\circ$493, is clear.}
\label{comp}
\end{figure}

Figure~\ref{mg_comp} shows a portion of the NUV spectra of \bd{}, in the
region of the \ion{Mg}{2} doublet at 2800\,{\AA}. \textit{HST}/STIS
spectra of two metal-poor giants with similar atmospheric parameters are
shown for comparison, \mbox{HD~108317} \cite[\teff = 5100~K,
\metal = $-$2.53; ][]{roederer2012d} and \mbox{HD~196944}
\citep[\teff = 5170~K, \metal = $-$2.46; ][]{roederer2008}. The effective
temperatures and surface gravities of \mbox{HD~108317} and
\mbox{HD~196944} are comparable to \bd, but their metallicities are
higher by about 1.5~dex. The lower metallicity of \bd\ is immediately
apparent from inspection of this figure. \mbox{HD~108317} is moderately
enhanced in $r$-process material, ([Eu/Fe]~$= +$0.5;
\citealt{roederer2012d}), and \mbox{HD~196944} is enhanced in $s$-process material
([Ba/Fe]~$= +$1.5; \citealt{roederer2014b}). \bd\ does not exhibit
neutron-capture element enhancements. To better illustrate the rather
striking differences, Figure~\ref{comp} shows portions of the NUV
spectra around the lines of several neutron-capture elements for the
same three stars shown in Figure~\ref{mg_comp}. Note in particular the
absence of absorption by \ion{Cd}{1}, \ion{Os}{2}, \ion{Lu}{2}, and
\ion{Pb}{1} for \bd{}, which are well-known abundance markers for the
operation of the $s$-process \citep{gallino1998,arlandini1999,sneden2008}.

\begin{figure*}[!ht]
\epsscale{1.15}
\plotone{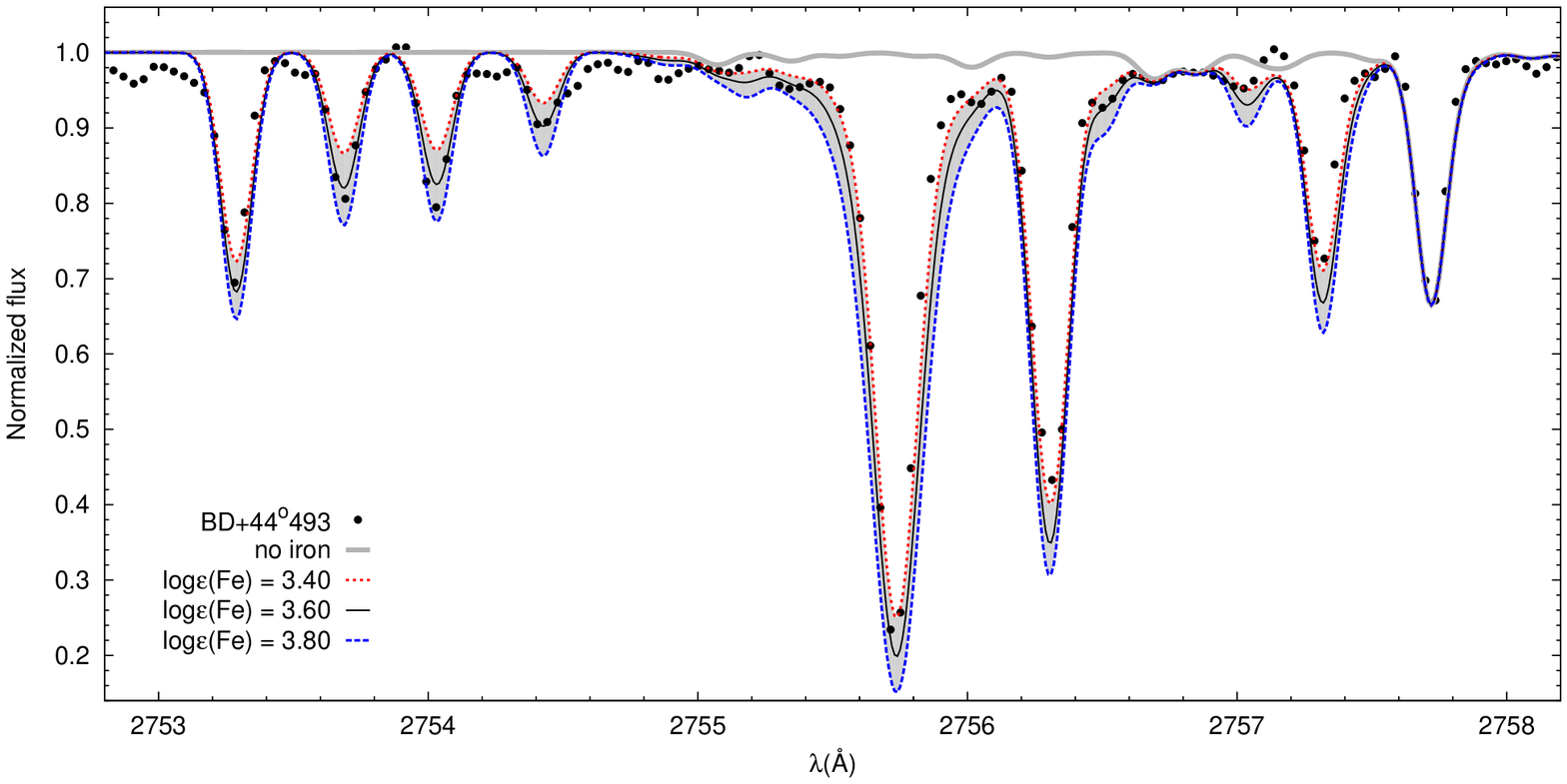}
\caption{Spectral synthesis of \ion{Fe}{1} and \ion{Fe}{2} features. 
The dots represent the observed spectra, the solid line
is the best abundance fit, and the dotted and dashed line are the lower
and upper abundance limits, indicating the abundance uncertainty. The shaded
area encompasses a 0.4~dex difference in \eps{Fe}.  The light gray line
shows the synthesized spectrum in the absence of Fe.}
\label{iron}
\end{figure*}

\subsection{Line Measurements}

Equivalent widths were obtained by fitting Gaussian profiles to the
observed atomic lines, using the {\tt Robospect} package
\citep{robospect}. The line lists were based on the compilation of
\citet{roederer2012d}, as well as on data retrieved from the VALD
database \citep{vald} and the National Institute of Standards and
Technology Atomic Spectra Database \citep[NIST; ][]{nist}. Abundances
for individual \ion{Fe}{1} and \ion{Fe}{2} lines, derived from
equivalent widths as well as from spectral synthesis, are listed in
Table~\ref{eqw}. Figure~\ref{iron} shows a sample of the NUV spectra,
with a number of the Fe lines used for the synthesis.

The abundances of all \ion{Fe}{1} and \ion{Fe}{2} lines in our STIS
spectrum were verified by spectral synthesis. We adopt the
model-atmosphere parameters derived by \citet{ito2013}, \teff = 5430~K,
\logg = 3.4 (cgs), $v_\mathrm{micro}$ = 1.3 km\,s$^{-1}$, and
\metal\ = $-3.8$. The iron abundances we derive from NUV lines
[\ion{Fe}{1}:\ log($\epsilon$) = 3.62~$\pm$~0.02; \ion{Fe}{2}:\
log($\epsilon$) = 3.63~$\pm$~0.02] differ little from the values
derived by \citeauthor{ito2013}\ from optical lines [\ion{Fe}{1}:\
log($\epsilon$) = 3.67~$\pm$~0.01; \ion{Fe}{2}:\ log($\epsilon$) =
3.68~$\pm$~0.03].

Previous work has shown that small differences between optical and NUV
\ion{Fe}{1} and \ion{Fe}{2} lines may exist \citep{roederer2010,roederer2012d}.
For example, Figure~6 of \citeauthor{ito2013}\ reveals a small ``dip''
in the abundances of \ion{Fe}{1} and \ion{Fe}{2} lines blueward of the
Balmer series limit in \bd. \citet{roederer2012d}, \citet{lawler2013},
and \citet{wood2013,wood2014} investigated several causes of this effect
for other metal-poor giants, but the differences are not fully
understood at present. To minimize this effect, we reference abundance
ratios of other elements derived from NUV transitions to the iron
abundance also derived from NUV transitions.

\begin{figure}[!ht]
\epsscale{1.15}
\plotone{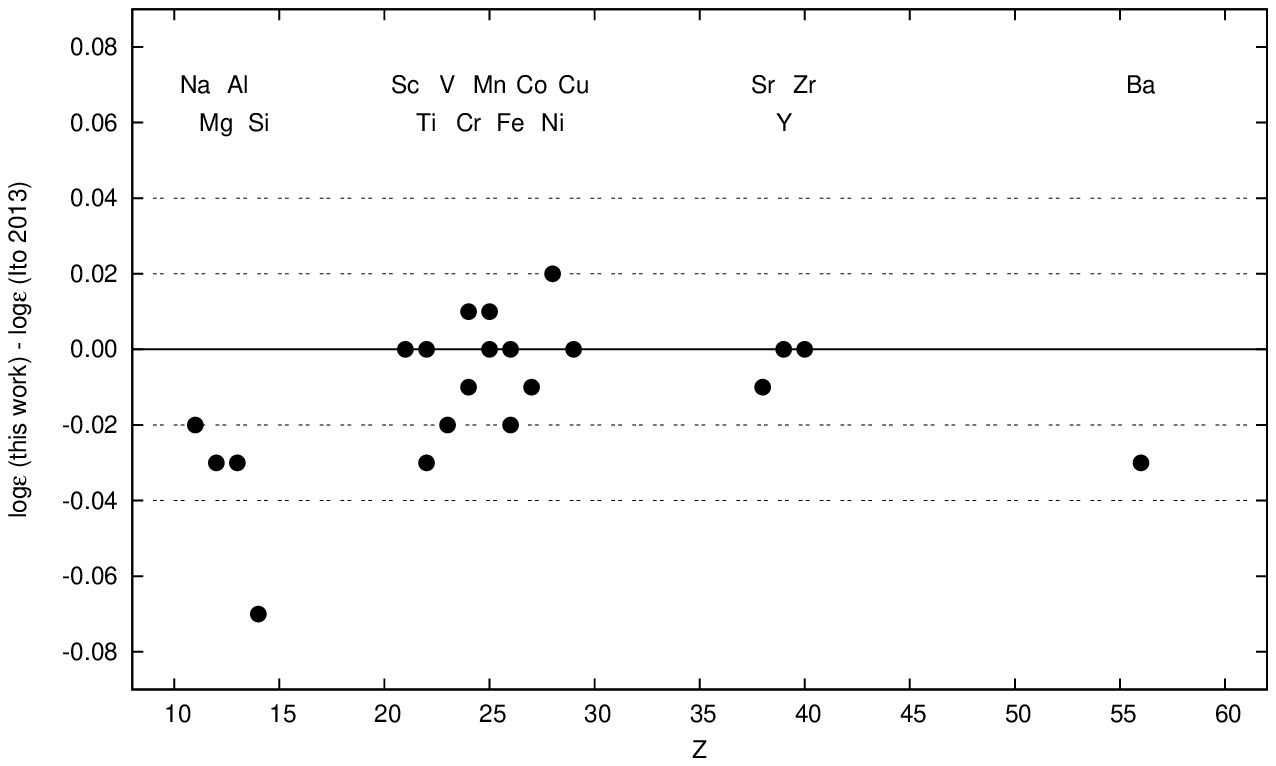}
\caption{Comparison between abundances determined from the equivalent-width analysis
of \citet{ito2013} and this work.  The agreement is quite satisfactory.}
\label{ito_comp}
\end{figure}

As a check on our procedures, we also used the equivalent-width values
published in \citet{ito2013} as input to our machinery.
Figure~\ref{ito_comp} shows the differences between the abundances
derived by equivalent-width analysis between \citeauthor{ito2013}\ and
this work. Apart from Si (which differs by $-$0.07~dex relative to Ito
et al.), all of the other differences lie within $\pm 0.04$~dex, with a
mean difference of $-$0.01~dex. This test demonstrates that our analysis
procedures and machinery can reproduce the \citeauthor{ito2013}\ values
to excellent precision.

\section{Abundance Analysis and Upper Limits}
\label{absec}

Chemical abundances or upper limits were obtained from the NUV spectrum
of \bd{} for 26 elements, including measurements for C, O, Sc, Ti, Cr,
Mn, Fe, and Ni, and upper limits for Be, B, Ge, Zr, Nb, Mo, Cd, Te, Ce, 
Nd, Eu, Gd, Yb, Lu, Hf, Os, Pt, and Pb.

\begin{figure*}[!ht]
\epsscale{1.15}
\plotone{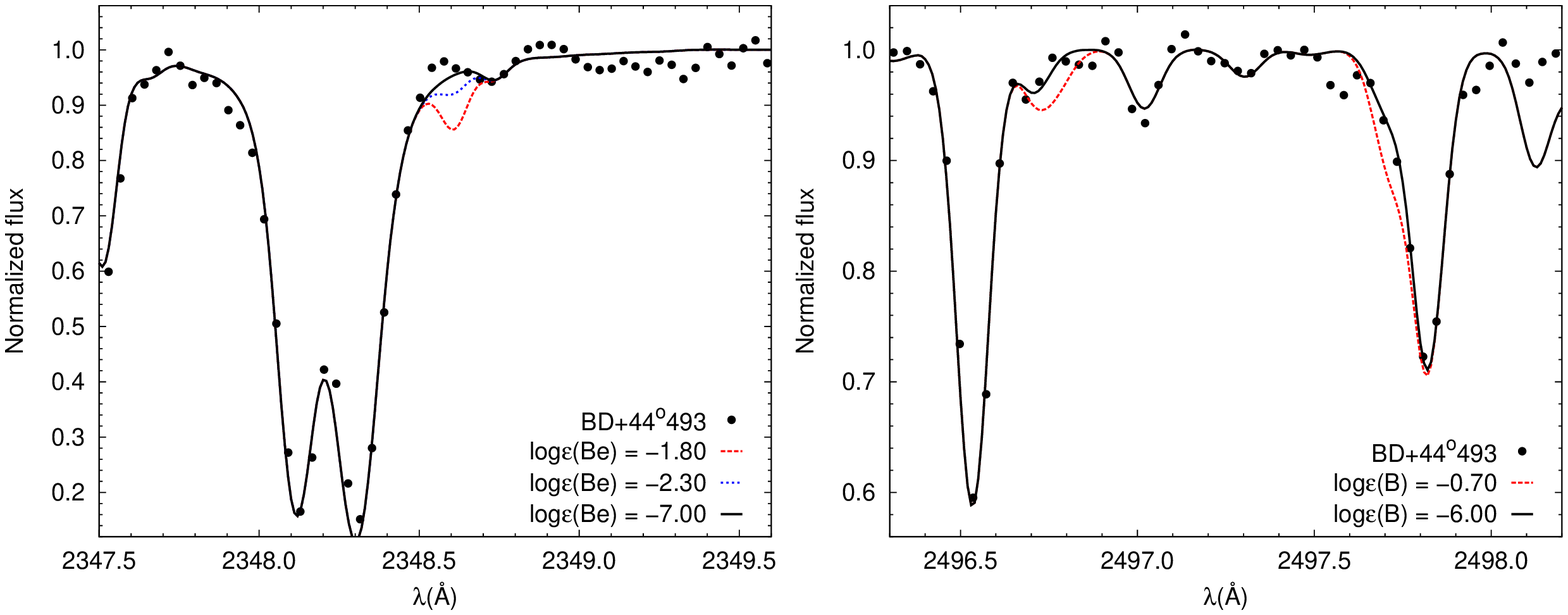}
\caption{Left panel: The \ion{Be}{1} line at 2348\,{\AA}, showing the 3-$\sigma$
upper limit from this work (\eps{Be} = $-2.30$) and from \citet{ito2013}
(\eps{Be} = $-1.80$). Right panel: The \ion{B}{1} line at 2497\,{\AA}, showing 
the 3-$\sigma$ upper limit from this work (\eps{B} = $-0.70$). 
See text for discussion.}
\label{beb_ul}
\end{figure*}

Our abundance analysis utilizes one-dimensional plane-parallel ATLAS9
model atmospheres with no overshooting \citep{castelli2004}, computed
under the assumption of local thermodynamic equilibrium (LTE). We use
the 2011 version of the MOOG synthesis code \citep{sneden1973} for this
analysis. To treat isotropic, coherent scattering in this version of
MOOG, the solution of the radiative transfer considers both absorption
and scattering components, rather than treating such scattering as pure
absorption \citep[see][for further details]{sobeck2011}.

Our final abundance ratios, [X/Fe], are given with respect to the solar
abundances of \citet{asplund2009}. Upper limits for elements for which
no absorption lines were detected provide additional information for the
interpretation of the overall abundance pattern of the stars. Based on
the S/N ratio in the spectral region of the line, and employing the
formula given in \citet{frebel2006b}, we derive 3-$\sigma$ upper
limits for 13 elements. Abundances and upper limits for individual
lines, derived from both equivalent widths and spectral synthesis, are
listed in Table ~\ref{abtable}.

A summary of the chemical abundances and upper limits for \bd{} is
provided in Table~\ref{abfinal}. The $\sigma$ refers to the standard
error of the mean. We have also addressed the systematic uncertainties that
could affect the model-atmosphere parameters. Table~\ref{sys} shows the
effect of changes in each atmospheric parameter on the determined
abundances, using spectral lines from which abundances were determined
by equivalent-width analysis alone. The adopted variations are 150~K for
\teff, 0.5~dex for \logg, and 0.3 km\,s$^{-1}$ for $v_\mathrm{micro}$. 
Also shown is the total uncertainty, taken as the quadratic sum of the
individual errors.

We discuss the determinations of beryllium, boron, carbon, the iron-peak
elements, and the neutron-capture elements in more detail in the
following subsections.

\subsection{Beryllium and Boron}

The \ion{Be}{1} resonance line at 2348\,{\AA} is not detected in our
spectrum of \bd. However, we can make use of this line to place a
significantly lower (by 0.5~dex, a factor of three) upper limit on the
Be abundance in \bd\ (\eps{Be}$ < -$2.3) than obtained by
\citet{ito2013} from the NUV \ion{Be}{2} doublet at 3130\,{\AA}
(\eps{Be}$ < -$1.8). The blue dotted line in the left panel of
Figure~\ref{beb_ul} shows our 3-$\sigma$ upper limit for the
\ion{Be}{1} 2348\,{\AA} line. For comparison, the red dashed line
shows the \citeauthor{ito2013}\ upper limit. We note, following
\citet{ito2013}, that 3D and NLTE effects on the beryllium abundance 
in metal-poor stars are expected to be small \citep{asplund2005}.

The \ion{B}{1} resonance doublet at 2497\,{\AA} is also not detected in
our spectrum of \bd. We derive an upper limit of \eps{B}~$< -$0.5 from
these lines, as shown in the right panel of Figure~\ref{beb_ul}. 
NLTE corrections for measured B abudances appear to be important
(e.g., \citealt{kiselman1996}), and could possibly perturb the results
(as inferred from the NLTE calculations of Kiselman \& Carlsson) by up to
$+$0.5 dex, for stars of metallicity similar to \bd. For the purpose of
comparing our present upper limit on B to previous detections, we prefer
to use the LTE results.

\begin{figure*}[!ht]
\epsscale{1.15}
\plotone{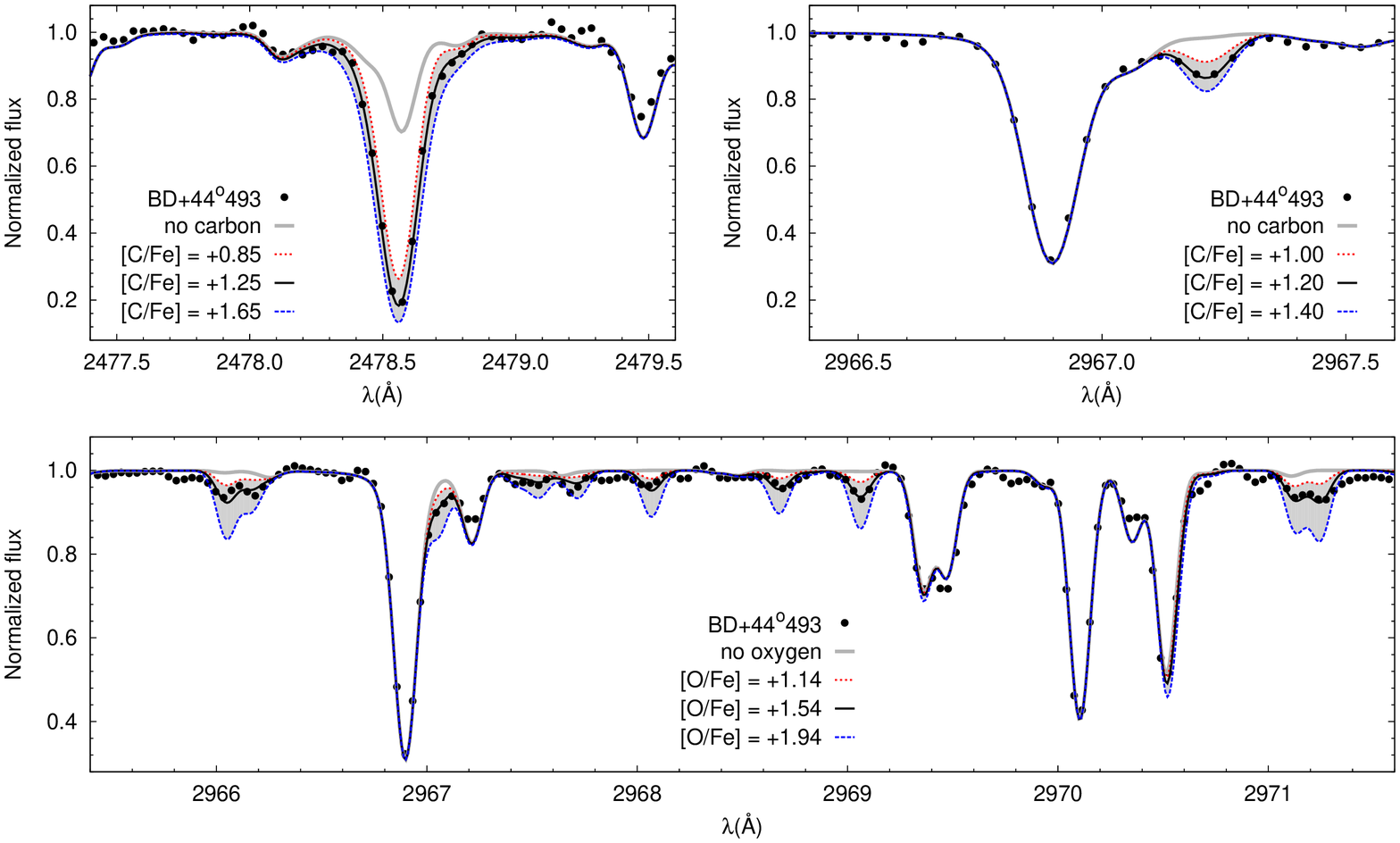}
\caption{Upper left panel: Spectral synthesis of the atomic \ion{C}{1}
feature at 2478.5\,{\AA}. 
The dots represent the observed spectra, the solid line is the best
abundance fit, and the dotted and dashed line are the lower and upper
abundance limits, indicating the abundance uncertainty. The shaded area
encompasses an 0.8~dex difference in [C/Fe]. The light gray line shows
the \ion{Fe}{2} line, located at essentially the same wavelength as the
\ion{C}{1} line. Upper right panel: \ion{C}{1} feature at 2967.2\,{\AA},
with the shaded area representing a 0.4~dex difference in [C/Fe]. Lower
panel: several OH features in the 2965-2972\,{\AA} range. The shaded
area shows a 0.8~dex difference in [O/Fe].}
\label{CO}
\end{figure*}

\subsection{Carbon and Oxygen}

The upper panels of Figure~\ref{CO} show the spectral synthesis of the
atomic \ion{C}{1} features at 2478.56\,{\AA} and 2967.21\,{\AA}.
Although the 2967.21\,{\AA} line is clean, the 2478\,{\AA} line is
blended with several other species, and in more metal-rich stars these
blends prohibit its use as an abundance indicator. The most severe of
these blending features is an \ion{Fe}{2} line at essentially the
identical wavelength, 2478.57\,\AA. The NIST ASD database reports
uncertainties of $\sigma \leq$ 18\% ($\leq$ 0.09 dex) on the $\log gf$
values of both the \ion{C}{1} lines. Thus, to the extent that we know
the Fe abundance, and are modeling the line formation appropriately, we
can use the \ion{C}{1} line as a C abundance indicator in \bd.

From the spectral synthesis of these lines, we obtain an average of
\eps{C} = 5.78, which yields a carbonicity of \cfe = $+$1.23 (using
[Fe/H] = $-$3.88). These values agree well with the optical
determinations from \citealt{ito2013} (\eps{C}= 5.95 and \cfe = $+$1.35,
using [Fe/H]=$-$3.83), but are a few tenths of a dex higher than the
near-infrared (NIR) determinations from \citealt{takeda2013} (\eps{C}=
5.69 and \cfe = $+$0.83, using [Fe/H] = $-$3.68). It must be kept in
mind that the optical values are determined from the CH $G$-band, and the
NIR from \ion{C}{1} $1.068-1.069~\mu$m lines.

\begin{figure*}[!ht]
\epsscale{1.15}
\plotone{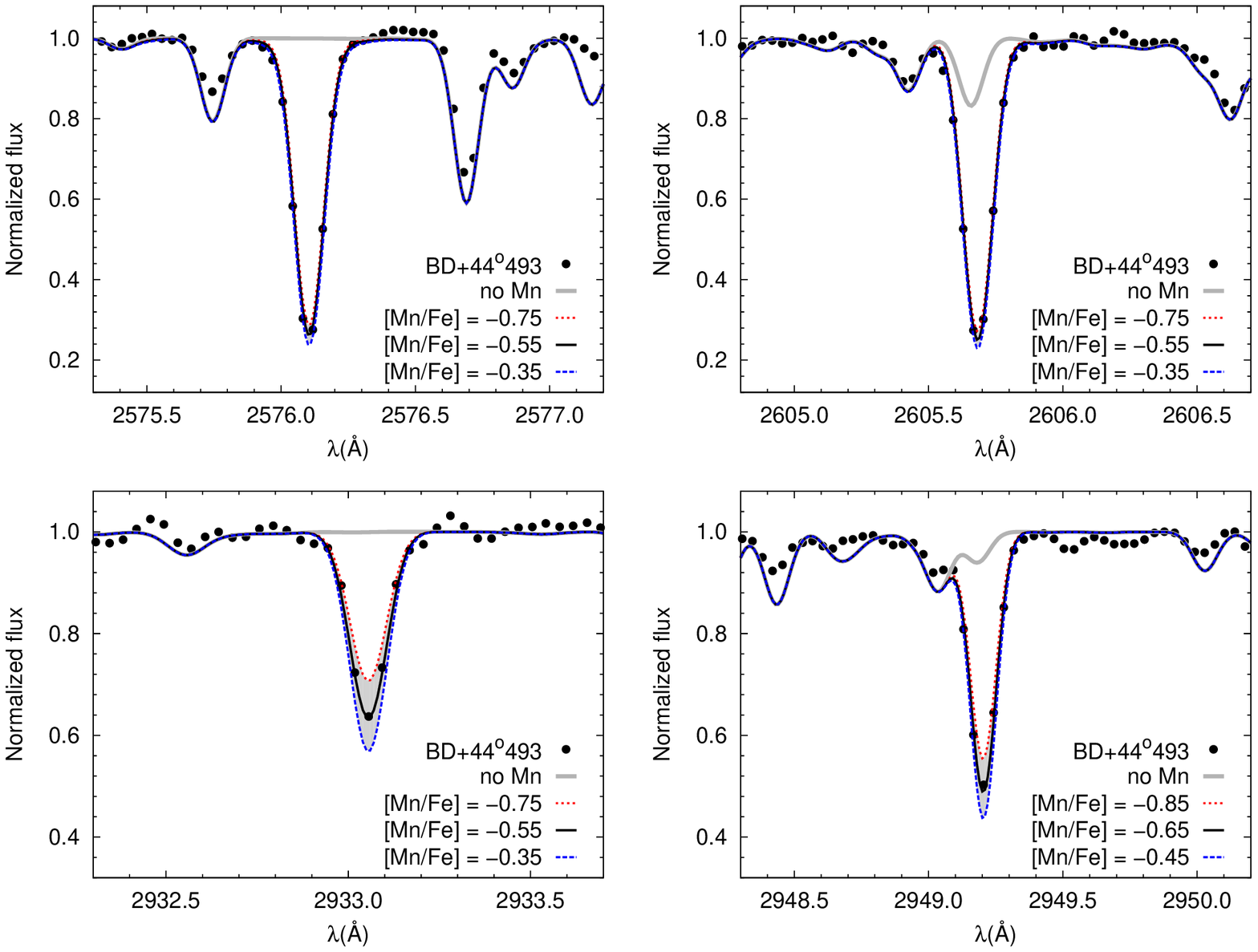}
\caption{Spectral synthesis of four \ion{Mn}{2} features. 
The dots represent the observed spectra, the solid line
is the best abundance fit, and the dotted and dashed line are the lower
and upper abundance limits, indicating the abundance uncertainty. The shaded
area encompasses an 0.6~dex difference in [Mn/Fe].  The light gray line
shows the synthesized spectrum in the absence of Mn.}
\label{mnii}
\end{figure*}

To our knowledge, this is the first determination of [C/Fe] based on
atomic lines in the NUV region. The relatively close agreement of this
determination with the [C/Fe] inferred from the molecular CH $G$-band is
encouraging. Previous modeling has suggested that 3D effects on the CH
and C$_2$ features for giants and subgiants at [Fe/H] $\sim -3.0$ can
lead to an over-estimate of [C/Fe] of $+0.5$ to $+0.8$ dex
\citep{asplund2005,collet2007}, although NLTE effects are not expected
to be large. Another possibility is that uncertainties in the UV 
opacity determination have the same
magnitude as 3D effects. \citet{schuler2008} found that the [C/Fe] ratio
derived from an LTE analysis of the [C I] forbidden line at 8727\,{\AA}
for the CEMP star HE~1005-1429 ([Fe/H] $= -3.08$) was on the order of
0.3-0.4 dex lower than the value reported by \citet{aoki2007}, based on
the molecular C$_2$ feature at 5170\,{\AA}. Future observations of the
NUV \ion{C}{1} features for additional (necessarily bright, and ideally
extremely metal-poor) CEMP stars may thus prove illuminating.

The lower panel of Figure~\ref{CO} shows the spectral region 2965\,
{\AA}--2972\,{\AA}, where several OH features are available. We were
able to obtain adequate fits for 11 lines with the same input abundance,
using the line list from \citet{kurucz1993}. The value of the O
abundance we obtain, [O/Fe] = $+$1.54, is in good agreement with that
derived by \citealt{ito2013} (also using an OH feature as an indicator),
[O/Fe] = $+$1.64.

\subsection{The Iron-Peak Elements}

Abundances for Sc, Ti, Cr, and Ni were determined with an
equivalent-width analysis only. \ion{Mn}{2} lines are broadened by
hyperfine splitting of the $^{55}$Mn isotope, so we derived those
abundances from spectral synthesis. Figure~\ref{mnii} shows the
synthesis of four \ion{Mn}{2} lines in the NUV spectrum of \bd.

To make a fair comparison with the \citet{ito2013} abundances, we
recomputed their optical abundances of Ti~\textsc{ii}, Mn~\textsc{ii},
and Ni~\textsc{i} on the same $\log gf$ scale we used for the NUV lines.
From the \citet{wood2013} study, we find that the Ito et al.\ Ti~\textsc{ii}
abundance would have decreased by only 0.01~dex. From lines in common
with the \citet{denhartog2011} study, we find that the Ito et al.\
Mn~\textsc{ii} abundance would have increased by 0.08~dex. From lines in
common with the \citet{wood2014} study, we find that the Ito et al.\
Ni~\textsc{i} abundance would have had no change. There would also be no
change for the Sc~\textsc{ii} abundance, since Ito et al.\ used the
$\log gf$ values reported by \citet{lawler1989}.

\begin{figure*}[!ht]
\epsscale{1.15}
\plotone{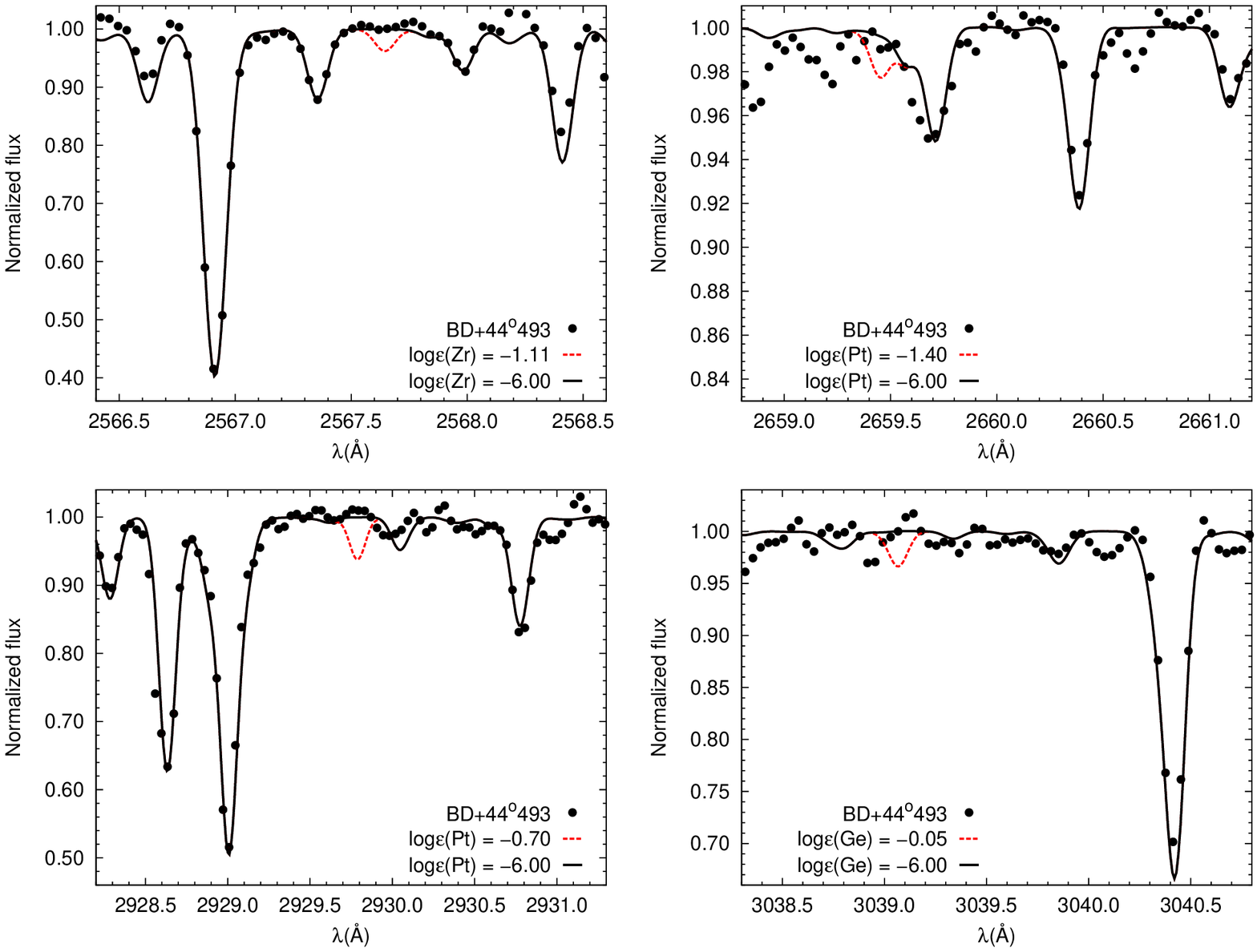}
\caption{Examples of the 3-$\sigma$ upper limits determined for \ion{Zr}{2},
\ion{Pt}{1}, and \ion{Ge}{1}.  The solid line shows the synthesized
spectrum in the absence of the labeled features.}
\label{upperlim}
\end{figure*}

As a result of this exercise, we find good agreement between the
abundance ratios determined by this work and those from \citet{ito2013}
-- these are \abund{\ion{Sc}{2}}{Fe} = $+$0.29, identical to that
obtained from the optical work ($+$0.29); \abund{\ion{Ti}{2}}{Fe} =
$+$0.36 ($+$0.36 in the optical); \abund{\ion{Cr}{2}}{Fe} = $-$0.09
($-$0.22 in the optical); \abund{\ion{Mn}{2}}{Fe} = $-$0.58 ($-$0.79 in
the optical); and \abund{\ion{Ni}{1}}{Fe} = $-$0.02 ($+$0.08 in the
optical).

\subsection{The Neutron-Capture Elements}

\begin{figure*}[!ht]
\epsscale{0.95}
\plotone{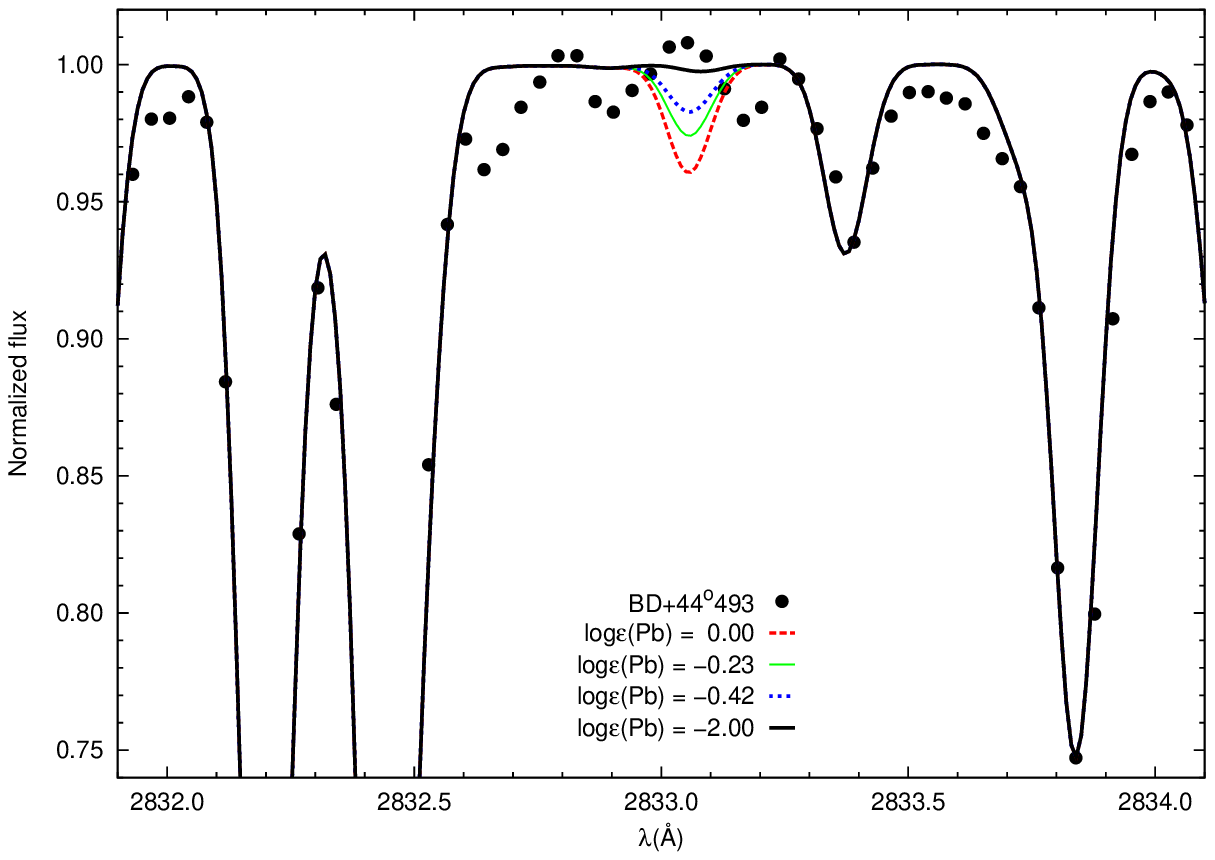}
\caption{The \ion{Pb}{1} line at 2833\,{\AA}, showing the 3-$\sigma$
(\eps{Pb} = $-0.23$) and 2-$\sigma$ (\eps{Pb} = $-0.42$) upper limits. See
text for discussion.}
\label{pb_ul}
\end{figure*}

Only upper limits were determined in \bd{} for neutron-capture elements
in the NUV region; Figure~\ref{upperlim} shows examples of these limits
for \ion{Zr}{2}, \ion{Ge}{1}, and \ion{Pt}{1}. Figure~\ref{pb_ul} shows
the comparison between the observed and the synthetic spectra around the
\ion{Pb}{1} 2833.05\,{\AA} feature. We determine a 3-$\sigma$
(2-$\sigma$) upper limit of \eps{Pb }$<-$0.23 ($< -$0.42) from this
line, assuming the poorly fit absorption features at 2832.9\,{\AA} and
2833.2\,{\AA} are noise, and not absorption lines. Our value confirms
and strengthens the \citet{ito2013} upper limit of \eps{Pb}$<-$0.10,
estimated from the weak optical \ion{Pb}{1} line at 4057.80\,{\AA}.
Syntheses of both the 3-$\sigma$ and 2-$\sigma$ upper limits are shown
in Figure~\ref{pb_ul}, as well as for \eps{Pb} = 0.00 and
\eps{Pb}=$-$2.00, for reference.

Figure~\ref{pb_ul} reveals that our upper limit on the abundance of Pb
might indeed be too conservative, depending on the nature of the nearby
noise features. The wavelength of the Pb~\textsc{i} line is known to
better than 1~m\AA\ \citep{wood1968}, so these supposed noise features
are not due to Pb~\textsc{i} absorption. The feature at 2832.9\,{\AA},
however, is also observed in our spectra of \mbox{HD~108317} and
\mbox{HD~196944} (see Figure~\ref{comp}), but the NIST database does not
include any probable lines at this wavelength. The known Fe~\textsc{ii}
line at 2833.09\,{\AA} \citep{roederer2012} does not appear in our
spectrum of \bd. The feature at 2833.2\,{\AA} may also appear in
\mbox{HD~108317} and \mbox{HD~196944}, but there it is weak. From
examination of the summed spectra from the first four observations and
the second four observations of \bd\ independently, these unidentified
features appear in both. This indicates that they are not random noise
spikes. If we do not treat these unfit features as noise, the upper
limit on Pb becomes tighter -- by at least several tenths of a dex --
but it is still limited by our ability to correctly identify the
continuum or other contaminants.

\section{Discussion and Conclusions}
\label{diss}

\begin{figure}[!ht]
\epsscale{1.15}
\plotone{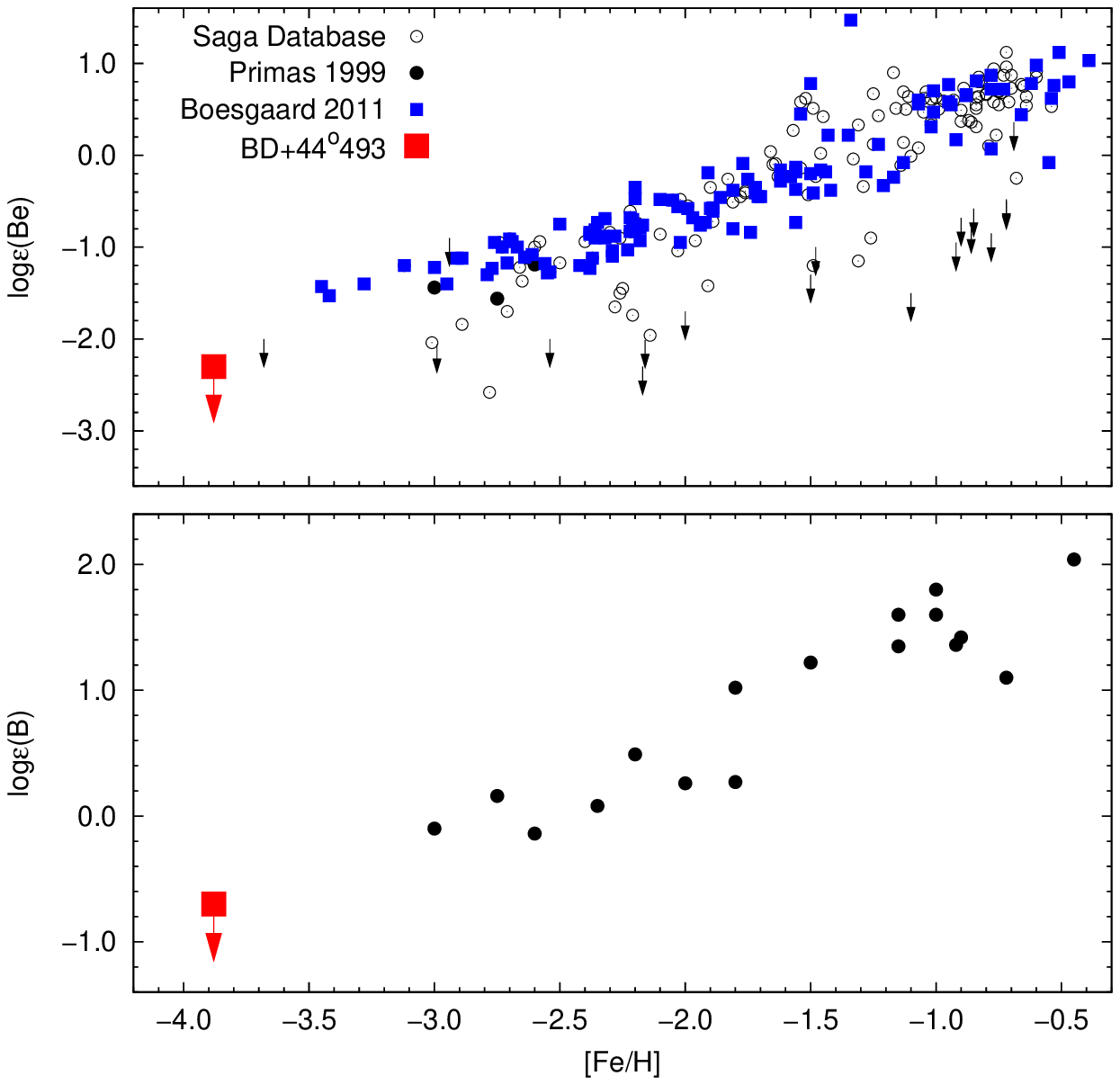}
\caption{Be and B abundances (and upper limits), as a function of the 
metallicity, for BD$+44^\circ493$ and  the stars listed in \citet{primas1999},
\citet{boesgaard2011}, and the SAGA database \citep{saga2008}. Individual 
references include: \citet{tan2009}, \citet{boesgaard2005}, \citet{boesgaard2006}, 
\citet{boesgaard2007}, \citet{garcia2006}, \citet{rich2009}, and 
\citet{smiljanic2009}.}
\label{bbefeh}
\end{figure}

\citet{ito2013} have discussed in detail the importance of the low upper limit
they derived for Be in the optical spectrum of \bd, which we have now
lowered by about a factor of three, to \eps{Be}$ < -$2.3, based on our
NUV \textit{HST}/STIS measurements. Our estimate of the upper limit on B
(\eps{B} $ < -0.7$), is also of significance, since these limits are at
the lowest level yet determined for very and extremely metal-poor stars.
Figure~\ref{bbefeh} shows a comparison between Be and B abundances, as a
function of \metal, for the upper limits determined in this work and
data from \citet{primas1999}, \citet{boesgaard2011}, and the SAGA
database \citep{saga2008}. Individual references are listed in the
caption of Figure~\ref{bbefeh}.

How can we account for the low upper limits for Be and B, taken together with
the fact that the Li abundance for \bd\ \citep[\eps{Li} = 1.0, reported
by][]{ito2013} is significantly below the level of the Spite Plateau?  The low
abundance of Li in this star, compared to the Spite Plateau value, could be a
result of convective mixing with internal layers in which Li is fully depleted.
Be might also be affected by mixing with material from layers that reach its
burning temperature ($3.0 \times 10^{6}$~K), which is slightly higher than that
of Li ($2.5 \times 10^{6}$~K). Compared to these two elements, B has an even
higher burning temperature ($5.0 \times 10^{6}$), and would be expected to be
less affected by mixing, although such an interpretation depends on detailed
modeling of stellar evolution during the subgiant phase.  Alternatively, the
very low upper limits of Be and B are consistent with the view that CEMP-no
stars such as \bd\ may have formed in the very early universe, {\it prior} to
the establishment of the level of cosmic-ray flux necessary to produce Be and B
by spallation (for a more detailed discussion, see the review by
\citealt{prantzos2012}).  The low abundance of Li could then be accounted for by
mixing of Li-astrated material (due to burning by first-generation stars) with
primordial Li created by Big Bang nucleosynthesis \citep{piau2006}.

This alternative receives some support from the recent observations of Li
abundances below the Spite Plateau for CEMP-no stars by \citet{hansen2014},
including two stars with \teff\ = 6100~K, presumably too warm for conventional
Li-depletion from convective mixing to have taken place. It is worth recalling
that \citet{masseron2012} reports that the CEMP-no class {\it only} contains
Li-depleted stars.  Unfortunately, Be and B abundance estimates are not yet
available for the CEMP-no stars of \citet{hansen2014}. Improved models and, in
particular, additional  observations of Be and B for CEMP-no stars, are
necessary in order to constrain these ideas further.  It should be noted that,
as already discussed by \citet{ito2013}, the progenitor of \bd{} is
unlikely to be a significant source of high-energy CNO nuclei that could yield
lighter elements by spallation processes. This fact might be a useful constraint
on the nature of the progenitor, most likely a faint (mixing and fallback)
supernova, that produces the high C and O abundances found in \bd.

We have discussed above that our measurement of [C/Fe], based on the NUV
atomic \ion{C}{1} lines at 2478.56\,{\AA} and 2967.21\,{\AA}, provides an
important validation of [C/Fe] estimates for CEMP stars based on the CH
$G$-band in the optical, as well as from \ion{C}{1} lines in the NIR.
Since this is the first determination of [C/Fe] from NUV spectroscopy,
it would clearly be important to carry out similar observations of
additonal bright CEMP stars, in order to test if this level of agreement
holds for stars that are cooler, or more metal-rich, than \bd.  

\begin{figure*}[!ht]
\epsscale{0.90}
\plotone{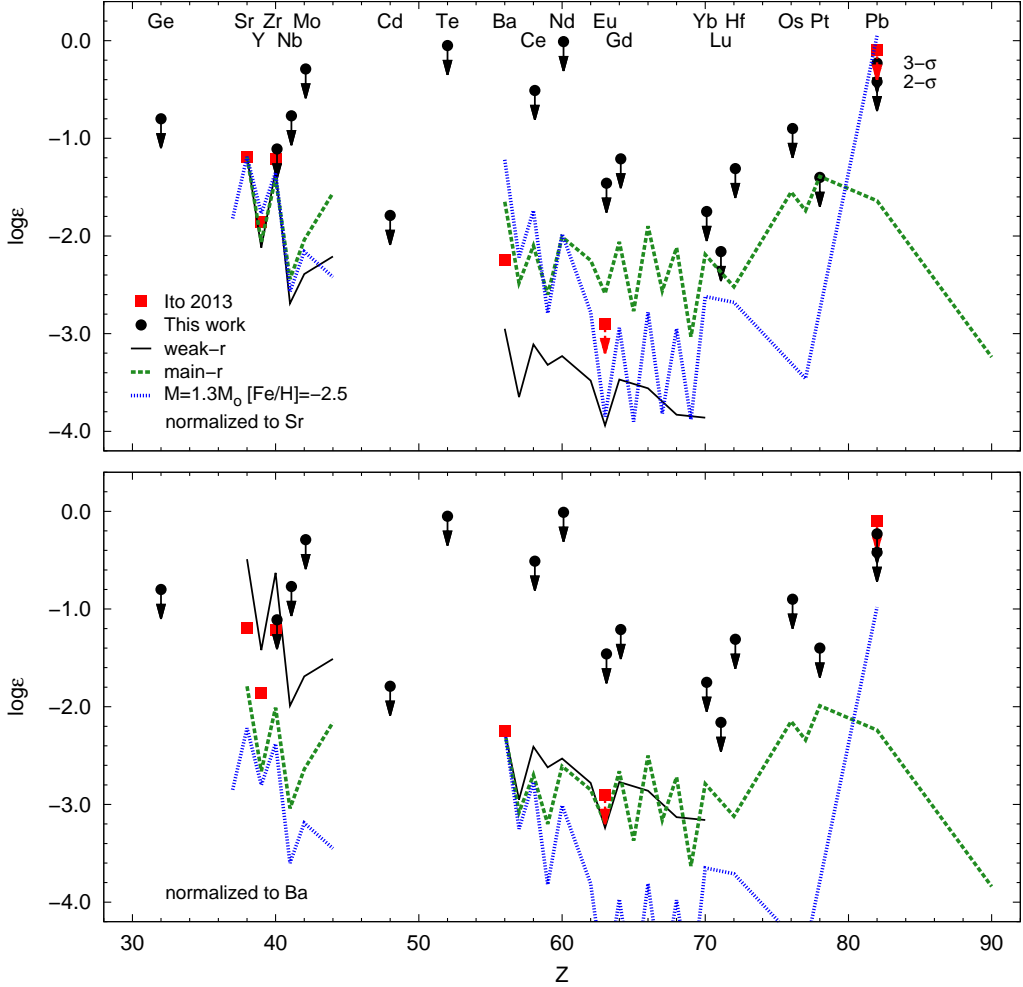}
\caption{Comparison between upper limits determined from this work, abundances
from \citet{ito2013}, and stellar templates for the solar $r$-process and the
1.3~M$_{\odot}$, \metal\ = $-$2.5 AGB yields described in \citet{placco2013}. Upper
panel: Models normalized to the Sr optical abundance. Lower panel: Models normalized
to the Ba optical abundance.}
\label{pattern}
\end{figure*}

\begin{figure*}[!ht]
\epsscale{1.00}
\plotone{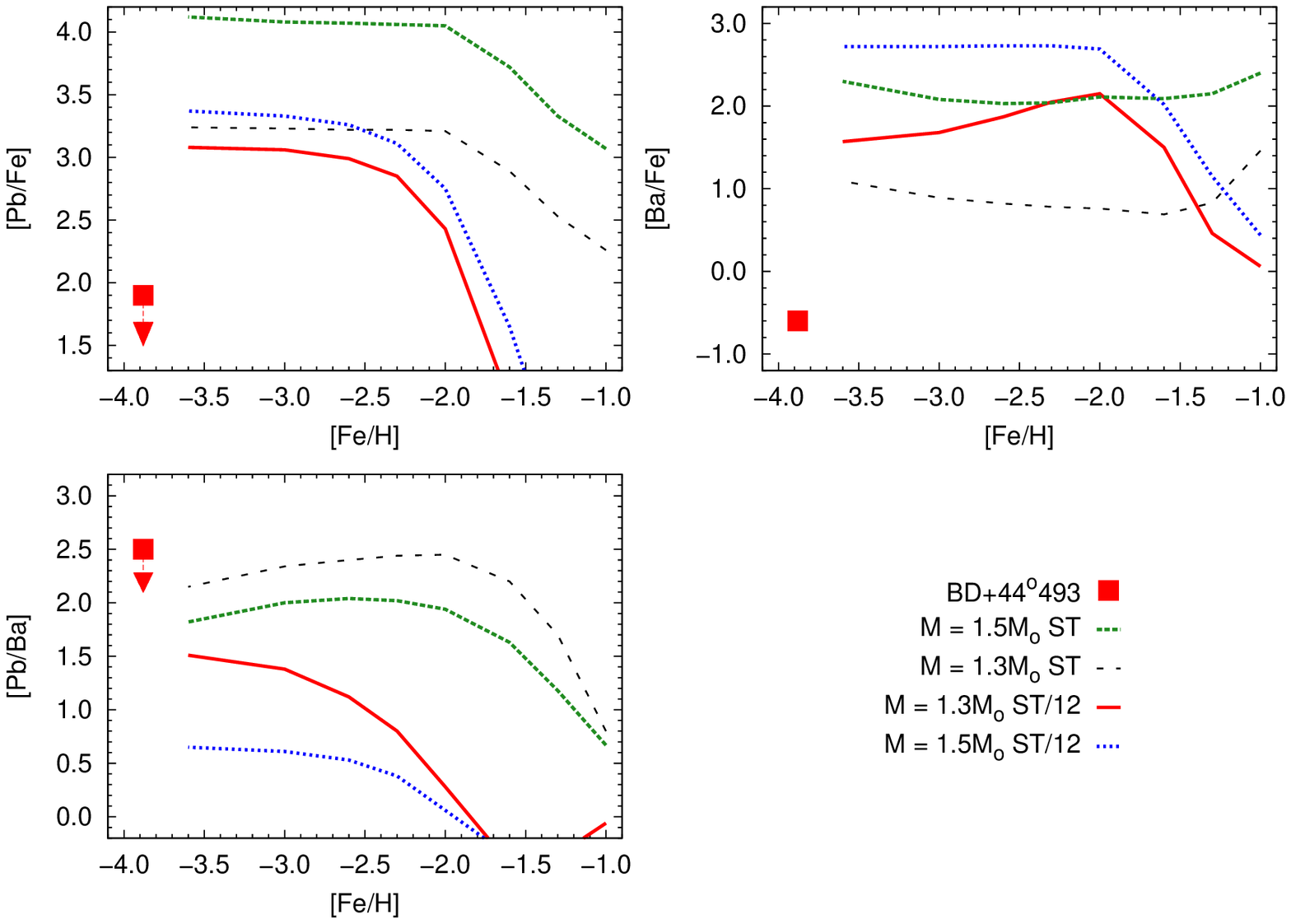}
\caption{\abund{Pb}{Fe} (this work), \abund{Ba}{Fe}, and \abund{Pb}{Ba}
\citep{ito2013} compared with AGB theoretical predictions for two different
initial masses and $^{13}$C-pocket efficiencies. Data were taken from Tables B5 
and B6 of \citet{bisterzo2010}.}
\label{bisterzo}
\end{figure*}

The \abund{Ni}{Fe} abundance ratio for \bd{} can also place further
constraints on Galactic chemical-evolution models. By comparing this
quantity with the value measured for the main-sequence turnoff star
HD~84937 \citep[\teff = 6300~K, \logg = 4.0, $v_\mathrm{micro}$ = 1.5
km\,s$^{-1}$, and \metal\ = $-2.32$;][]{wood2014}, one can see that, even
though their metallicities are more than 1.5~dex apart, their
\abund{Ni}{Fe} ratios exhibit only a 0.05~dex difference. This corroborates the
predicted plateau for Ni abundances as a function of the metallicity
from the theoretical models of \citet{kobayashi2011c}, and can set
observational limits on the Galactic initial mass function and yields
from both core-collapse supernovae and hypernovae.

Figure~\ref{pattern} shows a comparison between a set of $s$- and $r$-process
templates and the upper limits on neutron-capture elements determined in this
work. Also shown are the optical abundances from \citet{ito2013}. These are used
to normalize the models (to Sr in the upper panel and to Ba in the lower panel).
The template for the weak component of the $r$-process is the metal-poor giant
\mbox{HD~122563} \citep{honda2006,roederer2012d}.  The template for the main
component of the $r$-process is the metal-poor giant \mbox{CS~22892--052}
\citep{sneden2003,sneden2009,cowan2005, roederer2009}.  The template for the
$s$-process shows the AGB yields from the model presented in \citet{placco2013},
with M = 1.3~M$_{\odot}$ and [Fe/H] = $-$2.5\footnote{Intermediate-mass AGB
yields are beyond the scope of the current work, and were not added to this
analysis.}.  This is necessary due to the fact that, in contrast to the
``universal'' $r$-process, the $s$-process abundances are highly dependent on
metallicity.  These templates are not intended as firm representations of the
expected nucleosynthesis outcomes, since, for example, the yields depend on the
physical conditions at the time the nucleosynthesis events took place.

Stars with metallicities as low as \bd\ do not exhibit enhancements of
$s$-process material (e.g., \citealt{simmerer2004}), at least none have been
identified to date. At somewhat higher metallicities, the $s$-process elements
in CEMP-$s$ stars typically show substantial enhancements relative to the solar
abundance ratios (e.g., \citealt{aoki2002}), which are attributed to a
mass-transfer event from a companion star that passed through the AGB phase of
evolution. At lower metallicity, the high neutron-to-seed ratios are expected to
drive the flow to the most massive stable elements that can be produced by the
$s$-process, Pb and Bi (e.g., \citealt{gallino1998}). Enhanced Pb abundances are
considered unmistakable signatures of the operation of $s$-process
nucleosynthesis in a low-metallicity environment.

To the best of our knowledge, Pb has not been detected for any stars with
[Fe/H]~$< -$3.2 (see, e.g., Figure~3 of \citealt{roederer2010b}).  Our results
do not change this finding. Figure~\ref{bisterzo} shows the \abund{Pb}{Fe},
\abund{Ba}{Fe}, and \abund{Pb}{Ba} values for \bd, compared with four
different model prescriptions from \citet{bisterzo2010}. Even though \bd{}
exhibits a clear lack of Pb and Ba when compared to the models, the upper
limit we place on Pb, relative to barium, [Pb/Ba]~$< +$2.5, is on the cusp
of excluding $s$-process nucleosynthesis in low-mass low-metallicity AGB
stars with the highest neutron exposures possible.

This is not yet definitive evidence against an $s$-process origin of the
neutron-capture elements in \bd; however, the lack of significant
radial-velocity variations \citep[][spanning over 25 years]{carney2003,
ito2013} and the sub-solar [Sr/Fe] and [Ba/Fe] ratios are suggestive
that enrichment from an unseen companion that passed through the AGB
phase of evolution appears extremely unlikely, calling for non-AGB
sources, such as spinstars or faint supernovae explosions that undergo
mixing and fallback.

Some form of $r$-process nucleosynthesis may have been the dominant
production mechanism for the neutron-capture elements observed in \bd,
and other stars at extremely low metallicity (e.g., \citealt{truran1981}).
Barium is the heaviest element detected in \bd. The abundance pattern
presented in the bottom panel of Figure~\ref{pattern} cannot exclude
that the neutron-capture abundance pattern in \bd\ resembles that found
in the main or weak components of the $r$-process. \citet{roederer2014}
also reached this conclusion, based on their analysis of optical spectra
of \bd\ and other members of the CEMP-no sub-class of stars.

\bd\ is also the most metal-poor star where the Ge abundance can be reasonably
evaluated. The upper limit on the Ge abundance in \bd\ we obtain is only
about a factor of three higher than the mean [Ge/Fe] ratio found for stars with
$-$3.0~$<$~[Fe/H]~$< -$1.6 (see Figure~3 of \citealt{roederer2012c}). This
indicates that Ge was not manufactured in large quantities by the stars
that produced the metals in \bd. The primary nucleosynthesis mechanism
of Ge assumed for the stars with $-$3.0~$<$~[Fe/H]~$< -$1.6,
some form of explosive or charged-particle nucleosynthesis, is not
excluded as a possible source for whatever Ge may be present in \bd.

The Te upper limit we obtain is still about 2~dex higher than what might be
required to secure a Te detection in \bd, based on the Ba abundance in
\bd\ and the $\log \epsilon$~(Te/Ba) ratios reported by
\citet{roederer2012}. Similarly, the Cd upper limit is about 1~dex
higher than what might be required to secure a Cd detection in \bd. If
the 3$^{\rm rd}$ $r$-process peak elements in \bd\ are assumed to follow
a scaled-solar $r$-process abundance pattern, the Pt upper limit is
about 0.5~dex higher than what might be required to secure a Pt
detection in \bd. If the 3$^{\rm rd}$ $r$-process peak elements in \bd\
are deficient relative to Ba and the rare-earth elements, as found in
the metal-poor giants \mbox{HD~122563} and \mbox{HD~128279} by
\citet{roederer2012d}, it would be even more challenging to secure a Pt
detection in \bd. Te and Pt are expected to be two of the most abundant
elements heavier than Zr \citep{sneden2008}, yet it is unlikely that either of these
elements can be detected in \bd. This highlights the challenge of firmly
establishing the nucleosynthetic origins of the elements heavier than the
iron group for extremely metal-poor CEMP-no stars.

\acknowledgments 

V.M.P. acknowledges support from the Gemini Observatory.
T.C.B. and H.S. acknowledge partial support for this work from 
PHY 08-22648; Physics Frontier Center/{}Joint Institute or Nuclear
Astrophysics (JINA), awarded by the US National Science Foundation. 
A.F. is supported by NSF CAREER grant AST-1255160.
W.A. was supported by the JSPS Grants-in-Aid for Scientific Research (23224004).
Generous support for Program GO-12554 has been provided by 
NASA through a grant from the Space Telescope Science Institute,
which is operated by the Association of Universities for 
Research in Astronomy, Inc., under NASA contract NAS 5-26555.

\clearpage

\bibliographystyle{apj}

\clearpage

\LongTables

\begin{deluxetable}{cccccc} 
\tablewidth{0pc} 
\tablecaption{\label{eqw} Equivalent Width Measurements} 
\tablehead{ 
\colhead{$\lambda$}& 
\colhead{$\chi$} &  
\colhead{$\log\,gf$}& 
\colhead{$W$}&
\colhead{$\log\epsilon_W$}&
\colhead{$\log\epsilon_{syn}$}\\
\colhead{({\AA})}& 
\colhead{(eV)} &  
\colhead{}& 
\colhead{(m{\AA})}&
\colhead{}&
\colhead{}} 
\startdata 
\multicolumn{6}{c}{\ion{Fe}{1}} \\
\hline
2283.30 &  0.12 & $-$2.22 &   21.10 &    3.62 &    3.70 \\ 
2283.66 &  0.11 & $-$2.22 & \nodata & \nodata &    3.70 \\ 
2293.85 &  0.09 & $-$2.37 &   13.10 &    3.43 &    3.60 \\ 
2294.41 &  0.11 & $-$1.54 &   43.40 &    3.71 &    3.65 \\ 
2296.93 &  0.11 & $-$2.02 &   29.60 &    3.68 &    3.70 \\ 
2297.79 &  0.05 & $-$1.10 &   52.80 &    3.59 &    3.60 \\ 
2298.66 &  0.11 & $-$2.42 &   13.90 &    3.54 &    3.60 \\ 
2299.22 &  0.09 & $-$1.55 &   38.00 &    3.48 &    3.65 \\ 
2320.36 &  0.05 & $-$0.99 &   57.80 &    3.66 &    3.65 \\ 
2350.41 &  0.00 & $-$3.03 &    7.30 &    3.67 & \nodata \\ 
2369.46 &  0.11 & $-$2.19 &   19.10 &    3.49 &    3.50 \\ 
2371.43 &  0.09 & $-$1.95 &   33.30 &    3.67 &    3.70 \\ 
2374.52 &  0.12 & $-$2.10 &   22.90 &    3.53 &    3.60 \\ 
2389.97 &  0.09 & $-$1.57 &   39.70 &    3.50 &    3.60 \\ 
2443.87 &  0.86 & $-$1.24 &   28.00 &    3.58 &    3.65 \\ 
2445.21 &  0.86 & $-$2.02 &    5.40 &    3.40 & \nodata \\ 
2453.48 &  0.92 & $-$0.92 &   27.90 &    3.31 & \nodata \\ 
2457.60 &  0.86 & $-$0.32 &   48.00 &    3.32 & \nodata \\ 
2462.18 &  0.05 & $-$1.30 &   49.80 &    3.54 &    3.60 \\ 
2462.65 &  0.00 & $-$0.32 &   78.40 &    3.46 &    3.60 \\ 
2463.73 &  0.96 & $-$1.13 &   20.60 &    3.33 & \nodata \\ 
2468.88 &  0.86 & $-$0.62 &   42.50 &    3.41 &    3.55 \\ 
2470.97 &  0.92 & $-$1.62 &    8.40 &    3.27 & \nodata \\ 
2472.89 &  0.05 & $-$0.08 &   99.50 &    3.65 &    3.60 \\ 
2485.99 &  0.92 & $-$1.61 &   10.30 &    3.36 &    3.60 \\ 
2486.69 &  0.96 & $-$0.91 &   28.50 &    3.35 &    3.50 \\ 
2487.07 &  1.01 & $-$0.75 &   32.90 &    3.38 &    3.55 \\ 
2487.37 &  0.09 & $-$1.90 &   31.60 &    3.51 &    3.55 \\ 
2491.16 &  0.11 & $+$0.13 &  104.30 &    3.55 &    3.60 \\ 
2495.87 &  0.86 & $-$1.76 &   12.70 &    3.56 &    3.65 \\ 
2496.53 &  0.92 & $-$0.66 &   39.10 &    3.37 &    3.50 \\ 
2501.13 &  0.00 & $-$0.35 &   81.70 &    3.55 &    3.60 \\ 
2501.69 &  0.86 & $-$1.51 &   23.50 &    3.69 &    3.70 \\ 
2507.90 &  0.96 & $-$0.79 &   34.50 &    3.40 &    3.50 \\ 
2508.75 &  0.99 & $-$1.95 & \nodata & \nodata &    3.65 \\ 
2517.66 &  0.99 & $-$0.98 &   32.80 &    3.57 &    3.65 \\ 
2518.10 &  0.09 & $-$0.26 &   75.10 &    3.37 &    3.60 \\ 
2519.63 &  1.01 & $-$1.20 &   19.70 &    3.41 &    3.65 \\ 
2522.48 &  0.92 & $-$1.92 &   12.20 &    3.76 &    3.70 \\ 
2522.85 &  0.00 & $+$0.26 & \nodata & \nodata &    3.70 \\ 
2530.69 &  0.09 & $-$2.37 & \nodata & \nodata &    3.70 \\ 
2543.92 &  2.45 & $+$0.70 &   31.40 &    3.32 &    3.55 \\ 
2552.61 &  0.11 & $-$2.52 &   15.30 &    3.62 &    3.65 \\ 
2556.86 &  0.86 & $-$2.02 & \nodata & \nodata &    3.55 \\ 
2560.56 &  1.01 & $-$2.11 & \nodata & \nodata &    3.65 \\ 
2569.74 &  0.99 & $-$2.24 &    3.10 &    3.47 & \nodata \\ 
2576.69 &  0.86 & $-$0.91 &   35.90 &    3.43 &    3.55 \\ 
2584.54 &  0.86 & $-$0.39 &   51.00 &    3.42 &    3.60 \\ 
2610.75 &  0.09 & $-$2.96 &    5.10 &    3.45 &    3.70 \\ 
2612.77 &  0.05 & $-$2.59 &   17.30 &    3.68 &    3.70 \\ 
2618.02 &  0.96 & $-$0.97 &   38.60 &    3.66 &    3.65 \\ 
2618.71 &  0.00 & $-$2.43 &   14.20 &    3.35 &    3.50 \\ 
2623.37 &  0.11 & $-$2.57 & \nodata & \nodata &    3.65 \\ 
2623.53 &  0.96 & $-$0.70 &   48.40 &    3.72 &    3.60 \\ 
2632.24 &  0.99 & $-$1.20 &   25.60 &    3.54 &    3.60 \\ 
2632.59 &  0.09 & $-$2.33 &   23.80 &    3.67 &    3.65 \\ 
2635.81 &  0.99 & $-$0.81 &   43.60 &    3.69 &    3.60 \\ 
2636.48 &  0.92 & $-$2.04 &    6.00 &    3.49 &    3.55 \\ 
2641.03 &  2.45 & $-$1.25 &    1.70 &    3.72 & \nodata \\ 
2644.00 &  1.01 & $-$0.91 &   29.50 &    3.38 &    3.50 \\ 
2647.56 &  0.05 & $-$2.42 &   18.50 &    3.55 &    3.60 \\ 
2651.71 &  0.96 & $-$2.04 &    5.80 &    3.51 &    3.70 \\ 
2656.14 &  2.40 & $-$0.59 &    3.70 &    3.36 &    3.60 \\ 
2656.79 &  1.49 & $-$1.77 &    3.90 &    3.61 &    3.65 \\ 
2660.40 &  0.99 & $-$2.33 &    5.40 &    3.80 &    3.70 \\ 
2662.06 &  0.96 & $-$1.61 &    9.00 &    3.30 & \nodata \\ 
2679.06 &  0.86 & $-$0.75 &   50.90 &    3.72 &    3.60 \\ 
2680.45 &  0.99 & $-$1.74 &    7.40 &    3.35 &    3.50 \\ 
2689.21 &  0.92 & $-$0.89 &   41.70 &    3.61 &    3.60 \\ 
2690.07 &  0.00 & $-$2.72 &   11.00 &    3.48 &    3.55 \\ 
2699.11 &  0.92 & $-$1.26 &   24.50 &    3.47 &    3.60 \\ 
2710.54 &  1.61 & $-$1.33 &    4.30 &    3.34 &    3.60 \\ 
2714.87 &  0.96 & $-$2.19 &    3.50 &    3.41 &    3.60 \\ 
2723.58 &  0.09 & $-$0.72 &   73.50 &    3.66 &    3.70 \\ 
2726.05 &  1.01 & $-$1.21 &   19.60 &    3.37 &    3.55 \\ 
2728.02 &  0.92 & $-$1.46 &   19.10 &    3.50 &    3.55 \\ 
2735.47 &  0.92 & $-$0.40 &   54.50 &    3.53 & \nodata \\ 
2737.31 &  0.11 & $-$0.61 & \nodata & \nodata &    3.70 \\ 
2744.07 &  0.12 & $-$0.98 & \nodata & \nodata &    3.65 \\ 
2754.03 &  0.99 & $-$1.38 & \nodata & \nodata &    3.60 \\ 
2755.18 &  2.43 & $-$1.28 & \nodata & \nodata &    3.65 \\ 
2756.27 &  0.05 & $-$2.17 & \nodata & \nodata &    3.65 \\ 
2756.33 &  0.11 & $-$1.09 & \nodata & \nodata &    3.65 \\ 
2759.81 &  1.01 & $-$1.58 &   12.00 &    3.45 &    3.55 \\ 
2772.07 &  0.86 & $-$1.53 & \nodata & \nodata &    3.65 \\ 
2772.11 &  0.09 & $-$1.48 & \nodata & \nodata &    3.65 \\ 
2813.29 &  0.92 & $-$0.35 & \nodata & \nodata &    3.60 \\ 
2823.28 &  0.96 & $-$0.90 & \nodata & \nodata &    3.60 \\ 
2827.89 &  0.05 & $-$2.80 &    8.00 &    3.43 &    3.60 \\ 
2838.12 &  0.99 & $-$1.11 & \nodata & \nodata &    3.55 \\ 
2936.90 &  0.00 & $-$0.79 & \nodata & \nodata &    3.50 \\ 
2959.99 &  2.69 & $-$0.07 &    9.90 &    3.55 &    3.55 \\ 
2965.25 &  0.12 & $-$1.34 & \nodata & \nodata &    3.70 \\ 
2970.10 &  0.11 & $-$1.15 & \nodata & \nodata &    3.65 \\ 
2970.12 &  0.09 & $-$1.87 & \nodata & \nodata &    3.65 \\ 
2983.57 &  0.00 & $-$0.58 & \nodata & \nodata &    3.65 \\ 
2994.43 &  0.05 & $-$0.53 & \nodata & \nodata &    3.70 \\ 
2994.50 &  0.12 & $-$2.22 & \nodata & \nodata &    3.70 \\ 
2999.51 &  0.86 & $-$0.60 & \nodata & \nodata &    3.70 \\ 
3000.45 &  1.49 & $-$1.09 & \nodata & \nodata &    3.50 \\ 
3008.14 &  0.11 & $-$0.84 & \nodata & \nodata &    3.60 \\ 
3016.18 &  0.99 & $-$1.44 & \nodata & \nodata &    3.60 \\ 
3021.07 &  0.05 & $-$0.36 & \nodata & \nodata &    3.60 \\ 
3026.46 &  0.99 & $-$1.12 & \nodata & \nodata &    3.60 \\ 
3037.39 &  0.11 & $-$0.70 & \nodata & \nodata &    3.60 \\ 
3042.66 &  0.99 & $-$1.30 & \nodata & \nodata &    3.60 \\ 
3047.61 &  0.09 & $-$0.56 &   79.70 &    3.45 &    3.60 \\ 
3059.09 &  0.05 & $-$0.69 & \nodata & \nodata &    3.70 \\ 
\hline
\multicolumn{6}{c}{\ion{Fe}{2}} \\
\hline
2331.31 &  0.23 & $-$0.68 &  110.40 &    3.58 &    3.70 \\ 
2354.89 &  0.35 & $-$1.05 &   84.70 &    3.49 &    3.65 \\ 
2359.11 &  0.11 & $-$0.60 & \nodata & \nodata &    3.70 \\ 
2359.60 &  2.68 & $-$0.73 &   21.40 &    3.57 &    3.55 \\ 
2360.00 &  0.23 & $-$0.52 & \nodata & \nodata &    3.70 \\ 
2360.29 &  0.30 & $-$0.51 & \nodata & \nodata &    3.70 \\ 
2361.73 &  2.69 & $-$0.79 &   12.30 &    3.31 &    3.50 \\ 
2368.60 &  0.35 & $-$0.69 &  102.70 &    3.58 &    3.55 \\ 
2370.50 &  0.39 & $-$1.23 &   76.10 &    3.68 &    3.70 \\ 
2384.39 &  0.39 & $-$0.96 &   85.30 &    3.60 &    3.70 \\ 
2399.24 &  0.08 & $-$0.14 & \nodata & \nodata &    3.65 \\ 
2422.69 &  3.89 & $+$0.01 &    7.80 &    3.45 &    3.60 \\ 
2429.39 &  2.70 & $-$0.61 &   26.70 &    3.63 &    3.60 \\ 
2430.08 &  2.83 & $+$0.23 &   40.50 &    3.37 & \nodata \\ 
2432.87 &  4.08 & $+$0.55 &   12.90 &    3.36 &    3.55 \\ 
2433.50 &  2.68 & $-$0.86 &   23.80 &    3.76 &    3.70 \\ 
2446.47 &  2.66 & $-$0.42 &   31.30 &    3.54 &    3.60 \\ 
2447.20 &  3.89 & $-$0.21 &    8.50 &    3.71 &    3.65 \\ 
2463.28 &  3.15 & $-$0.19 &   23.20 &    3.54 &    3.60 \\ 
2464.91 &  3.23 & $-$0.09 &   22.00 &    3.48 &    3.60 \\ 
2468.30 &  2.68 & $-$1.05 &   15.90 &    3.67 &    3.70 \\ 
2470.41 &  3.23 & $-$0.48 &   11.20 &    3.47 &    3.50 \\ 
2470.67 &  2.83 & $-$0.07 &   40.50 &    3.65 &    3.60 \\ 
2476.27 &  2.69 & $-$1.05 &   12.70 &    3.56 &    3.60 \\ 
2497.82 &  3.23 & $-$0.03 &   23.10 &    3.44 &    3.60 \\ 
2502.39 &  3.22 & $+$0.03 &   23.80 &    3.40 &    3.55 \\ 
2506.09 &  3.20 & $-$0.03 &   20.10 &    3.32 &    3.55 \\ 
2527.10 &  2.66 & $-$0.45 &   32.10 &    3.56 &    3.70 \\ 
2529.55 &  2.81 & $+$0.33 &   51.90 &    3.64 &    3.70 \\ 
2533.63 &  2.66 & $+$0.35 &   48.60 &    3.33 & \nodata \\ 
2555.07 &  2.84 & $-$0.81 &   12.30 &    3.44 &    3.60 \\ 
2555.45 &  2.86 & $-$0.83 &   11.30 &    3.43 &    3.60 \\ 
2559.77 &  3.23 & $-$0.72 &    9.30 &    3.59 &    3.60 \\ 
2560.28 &  3.20 & $-$0.16 & \nodata & \nodata &    3.60 \\ 
2562.54 &  0.99 & $+$0.02 &  105.30 &    3.50 &    3.65 \\ 
2563.48 &  1.04 & $-$0.23 &   94.60 &    3.64 &    3.70 \\ 
2566.62 &  2.81 & $-$1.07 &    6.60 &    3.34 &    3.55 \\ 
2566.91 &  1.08 & $-$0.64 & \nodata & \nodata &    3.70 \\ 
2582.58 &  1.08 & $-$0.45 &   77.50 &    3.52 &    3.65 \\ 
2590.55 &  2.70 & $-$1.32 &    7.20 &    3.52 &    3.60 \\ 
2591.54 &  1.04 & $-$0.46 &   80.40 &    3.56 &    3.65 \\ 
2592.78 &  4.08 & $+$0.65 &   18.20 &    3.44 &    3.50 \\ 
2608.85 &  2.81 & $-$1.39 &    7.80 &    3.73 &    3.70 \\ 
2610.63 &  0.05 & $+$0.92 & \nodata & \nodata &    3.70 \\ 
2620.17 &  2.84 & $-$1.15 & \nodata & \nodata &    3.65 \\ 
2620.70 &  2.83 & $-$0.55 & \nodata & \nodata &    3.60 \\ 
2621.67 &  0.12 & $-$0.94 &  110.30 &    3.63 &    3.70 \\ 
2626.50 &  2.86 & $-$0.67 &   12.60 &    3.30 &    3.50 \\ 
2637.64 &  3.34 & $-$0.56 &    6.80 &    3.37 & \nodata \\ 
2652.57 &  3.27 & $-$1.43 &    2.50 &    3.69 & \nodata \\ 
2664.66 &  3.39 & $+$0.31 &   29.90 &    3.43 &    3.55 \\ 
2684.75 &  3.81 & $+$0.23 &   13.00 &    3.38 &    3.60 \\ 
2721.81 &  3.15 & $-$1.25 &    4.20 &    3.63 &    3.70 \\ 
2732.45 &  0.23 & $-$2.96 &   39.70 &    3.68 &    3.70 \\ 
2769.36 &  3.15 & $-$0.48 &   18.20 &    3.60 &    3.60 \\ 
2917.47 &  1.04 & $-$2.85 & \nodata & \nodata &    3.65 \\ 
2965.41 &  3.42 & $-$2.24 & \nodata & \nodata &    3.65 \\ 
\enddata
\end{deluxetable}

\begin{deluxetable}{lcccrr}
\tablecolumns{6} 
\tablewidth{0pc} 
\tablecaption{Abundances and Upper Limits\label{abtable}} 
\tablehead{ 
\colhead{Species}& 
\colhead{$\lambda$}& 
\colhead{$\chi$} &  
\colhead{$\log\,gf$}& 
\colhead{$\log\epsilon$\,(X)}&
\colhead{Ref.}\\
\colhead{}& 
\colhead{({\AA})}& 
\colhead{(eV)} &  
\colhead{}& 
\colhead{}& 
\colhead{}} 
\startdata 
\ion{Be}{1} & 2348.61 & 0.000 & $+$0.140 & $<-$2.30 &  1 \\
\ion{B}{1}  & 2496.77 & 0.000 & $-$0.800 & $<-$0.70 &  1 \\
\ion{B}{1}  & 2497.72 & 0.002 & $-$0.500 & $<-$0.70 &  1 \\
\ion{C}{1}  & 2478.56 & 2.682 & $-$1.110 & $+$5.80\tablenotemark{a}  &  1 \\
\ion{C}{1}  & 2967.21 & 0.005 & $-$6.800 & $+$5.75\tablenotemark{a}  &  1 \\
\ion{Sc}{2} & 2552.35 & 0.022 & $+$0.030 & $-$0.44  &  2 \\ 
\ion{Ti}{2} & 2524.64 & 0.122 & $-$1.320 & $+$1.42  &  3 \\ 
\ion{Ti}{2} & 2525.60 & 0.151 & $-$0.570 & $+$1.44  &  3 \\ 
\ion{Ti}{2} & 2531.25 & 0.135 & $-$0.670 & $+$1.48  &  3 \\ 
\ion{Ti}{2} & 2534.62 & 0.122 & $-$0.930 & $+$1.44  &  3 \\ 
\ion{Ti}{2} & 2841.93 & 0.607 & $-$0.590 & $+$1.35  &  3 \\ 
\ion{Ti}{2} & 2888.93 & 0.574 & $-$1.360 & $+$1.49  &  3 \\ 
\ion{Ti}{2} & 2891.06 & 0.607 & $-$1.140 & $+$1.44  &  3 \\ 
\ion{Ti}{2} & 3058.09 & 1.180 & $-$0.420 & $+$1.42  &  3 \\ 
\ion{Cr}{2} & 2740.10 & 1.506 & $-$1.090 & $+$1.62  &  4 \\ 
\ion{Cr}{2} & 2751.87 & 1.525 & $-$0.290 & $+$1.72  &  4 \\ 
\ion{Cr}{2} & 2757.72 & 1.506 & $-$0.360 & $+$1.66  &  4 \\ 
\ion{Cr}{2} & 2762.59 & 1.525 & $+$0.050 & $+$1.68  &  4 \\ 
\ion{Cr}{2} & 2766.54 & 1.549 & $+$0.320 & $+$1.66  &  4 \\ 
\ion{Mn}{2} & 2576.11 & 0.000 & $+$0.400 & $+$1.00\tablenotemark{a} &  5 \\ 
\ion{Mn}{2} & 2605.68 & 0.000 & $+$0.136 & $+$1.00\tablenotemark{a} &  5 \\ 
\ion{Mn}{2} & 2933.05 & 1.175 & $-$0.102 & $+$1.00\tablenotemark{a} &  5 \\ 
\ion{Mn}{2} & 2949.20 & 1.175 & $+$0.253 & $+$0.90\tablenotemark{a} &  5 \\ 
\ion{Ni}{1} & 2289.99 & 0.000 & $+$0.060 & $+$2.24  &  6 \\ 
\ion{Ni}{1} & 2293.12 & 0.109 & $-$0.970 & $+$2.25  &  6 \\ 
\ion{Ni}{1} & 2312.34 & 0.165 & $+$0.410 & $+$2.48  &  6 \\ 
\ion{Ni}{1} & 2325.80 & 0.165 & $+$0.400 & $+$2.23  &  6 \\ 
\ion{Ni}{1} & 2346.63 & 0.165 & $-$0.840 & $+$2.33  &  6 \\ 
\ion{Ni}{1} & 2356.87 & 0.025 & $-$1.510 & $+$2.23  &  6 \\ 
\ion{Ni}{1} & 2360.64 & 0.275 & $-$1.080 & $+$2.24  &  6 \\ 
\ion{Ni}{1} & 2376.02 & 0.109 & $-$1.700 & $+$2.36  &  6 \\ 
\ion{Ni}{1} & 2386.59 & 0.109 & $-$1.180 & $+$2.28  &  6 \\ 
\ion{Ni}{1} & 2419.31 & 0.165 & $-$1.050 & $+$2.27  &  6 \\ 
\ion{Ni}{1} & 2821.29 & 0.025 & $-$1.410 & $+$2.37  &  6 \\ 
\ion{Ni}{1} & 2943.91 & 0.025 & $-$1.170 & $+$2.34  &  6 \\ 
\ion{Ni}{1} & 2992.59 & 0.025 & $-$1.220 & $+$2.48  &  6 \\ 
\ion{Ni}{1} & 3003.62 & 0.109 & $-$0.320 & $+$2.17  &  6 \\ 
\ion{Ni}{1} & 3012.00 & 0.423 & $+$0.000 & $+$2.23  &  6 \\ 
\ion{Ni}{1} & 3031.87 & 0.000 & $-$1.810 & $+$2.46  &  6 \\ 
\ion{Ni}{1} & 3037.93 & 0.025 & $-$0.520 & $+$2.40  &  6 \\ 
\ion{Ni}{1} & 3050.82 & 0.025 & $-$0.100 & $+$2.33  &  6 \\ 
\ion{Ni}{1} & 3054.31 & 0.109 & $-$0.600 & $+$2.39  &  6 \\ 
\ion{Ni}{2} & 2278.77 & 1.680 & $+$0.190 & $+$2.21  &  7 \\ 
\ion{Ni}{2} & 2297.14 & 1.254 & $-$0.070 & $+$2.27  &  7 \\ 
\ion{Ni}{2} & 2297.49 & 1.322 & $-$0.330 & $+$2.19  &  7 \\ 
\ion{Ni}{2} & 2356.40 & 1.859 & $-$0.830 & $+$2.29  &  7 \\ 
\ion{Ni}{2} & 2387.76 & 1.680 & $-$1.070 & $+$2.32  &  7 \\ 
\ion{Ni}{2} & 2394.52 & 1.680 & $+$0.170 & $+$2.09  &  7 \\ 
\ion{Ni}{2} & 2416.14 & 1.859 & $+$0.130 & $+$2.20  &  7 \\ 
\ion{Ni}{2} & 2437.89 & 1.680 & $-$0.330 & $+$2.13  &  7 \\
\ion{Ge}{1} & 2651.17 & 0.175 & $+$0.020 & $<-$0.80 &  8 \\ 
\ion{Ge}{1} & 2691.34 & 0.069 & $-$0.700 & $<-$0.00 &  8 \\ 
\ion{Ge}{1} & 3039.07 & 0.883 & $+$0.070 & $<-$0.05 &  8 \\ 
\ion{Zr}{2} & 2567.64 & 0.000 & $-$0.170 & $<-$1.11 &  9 \\ 
\ion{Zr}{2} & 2699.60 & 0.039 & $-$0.660 & $<-$0.40 & 10 \\ 
\ion{Zr}{2} & 2700.14 & 0.095 & $-$0.080 & $<-$1.03 &  9 \\ 
\ion{Zr}{2} & 2732.72 & 0.095 & $-$0.490 & $<-$0.45 &  9 \\ 
\ion{Zr}{2} & 2758.81 & 0.000 & $-$0.560 & $<-$0.46 &  9 \\ 
\ion{Zr}{2} & 2915.99 & 0.466 & $-$0.500 & $<-$0.38 &  9 \\ 
\ion{Zr}{2} & 3054.84 & 1.010 & $+$0.080 & $<-$0.16 & 10 \\ 
\ion{Nb}{2} & 2950.88 & 0.510 & $+$0.240 & $<-$0.77 & 11 \\ 
\ion{Nb}{2} & 3028.44 & 0.440 & $-$0.200 & $<-$0.51 & 11 \\ 
\ion{Mo}{2} & 2871.51 & 1.538 & $+$0.060 & $<-$0.29 & 12 \\ 
\ion{Cd}{1} & 2288.02 & 0.000 & $+$0.150 & $<-$1.79 & 13 \\ 
\ion{Te}{1} & 2385.79 & 0.589 & $-$0.810 & $<-$0.05 & 14 \\ 
\ion{Ce}{2} & 3063.00 & 0.900 & $+$0.400 & $<-$0.51 & 15 \\ 
\ion{Nd}{2} & 3014.17 & 0.200 & $-$0.660 & $<-$0.01 & 16 \\ 
\ion{Eu}{2} & 2906.67 & 0.000 & $-$0.350 & $<-$1.46 & 16 \\ 
\ion{Gd}{2} & 3010.13 & 0.000 & $+$0.190 & $<-$1.21 & 17 \\ 
\ion{Yb}{2} & 2891.39 & 0.000 & $-$1.169 & $<-$1.75 & 16 \\ 
\ion{Lu}{2} & 2615.43 & 0.000 & $-$0.270 & $<-$2.16 & 18 \\  
\ion{Hf}{2} & 2641.41 & 1.036 & $+$0.570 & $<-$1.11 & 19 \\ 
\ion{Hf}{2} & 2820.23 & 0.380 & $-$0.140 & $<-$1.31 & 19 \\ 
\ion{Hf}{2} & 2929.64 & 0.000 & $-$0.940 & $<-$1.02 & 19 \\ 
\ion{Os}{1} & 3058.66 & 0.000 & $-$0.410 & $<-$0.30 & 20 \\ 
\ion{Os}{2} & 2282.28 & 0.000 & $-$0.050 & $<-$0.90 & 20 \\ 
\ion{Pt}{1} & 2659.45 & 0.000 & $-$0.030 & $<-$1.40 & 21 \\ 
\ion{Pt}{1} & 2929.79 & 0.000 & $-$0.700 & $<-$0.66 & 21 \\ 
\ion{Pb}{1} & 2833.05 & 0.000 & $-$0.500 & $<-$0.23 & 22    

\enddata
\tablenotetext{a}{Synthesis}
\tablerefs{
(1)  \citet{nist};
(2)  \citet{lawler1989};
(3)  \citet{wood2013};
(4)  \citet{bergeson1993};
(5)  \citet{denhartog2011};
(6)  \citet{wood2014};
(7)  \citet{fedchak1999};
(8)  \citet{fuhr2009};
(9)  \citet{ljung2006};
(10) \citet{malcheva2006};
(11) \citet{nilsson2008};
(12) \citet{sikstrom2001};
(13) \citet{morton2000};
(14) \citet{roederer2012};
(15) \citet{biemont1999};
(16) \citet{kurucz1995};
(17) \citet{denhartog2006};
(18) \citet{roederer2010b};
(19) \citet{lawler2007};
(20) \citet{quinet2006};
(21) \citet{denhartog2005};
(22) \citet{biemont2000}, using hfs presented in Appendix C of
     \citet{roederer2012d}.
}

\end{deluxetable}

\clearpage

\begin{deluxetable}{lcrrrr}
\tablecolumns{6} 
\tablewidth{0pc} 
\tablecaption{Final NUV LTE Abundances of BD+44$^\circ$493\label{abfinal}} 
\tablehead{ 
\colhead{Species}& 
\colhead{$\log\epsilon_{\odot}$\,(X)}&
\colhead{$\log\epsilon$\,(X)}&
\colhead{\mbox{[X/Fe]}}&
\colhead{$\sigma$}& 
\colhead{$N$}}
\startdata 
\ion{Be}{1}  & 1.38 & $<-$2.30 &                  $<+$0.20 & \nodata &   1 \\
\ion{B}{1}   & 2.70 & $<-$0.70 &                  $<+$0.48 & \nodata &   2 \\
\ion{C}{1}   & 8.43 &     5.78 &                   $+$1.23 &    0.20 &   2 \\
OH           & 8.69 &     6.35 &                   $+$1.54 &    0.20 &  11 \\
\ion{Sc}{2}  & 3.15 &  $-$0.44 &                   $+$0.29 &    0.10 &   1 \\
\ion{Ti}{2}  & 4.95 &     1.43 &                   $+$0.36 &    0.02 &   8 \\
\ion{Cr}{2}  & 5.64 &     1.67 &                   $-$0.09 &    0.02 &   5 \\
\ion{Mn}{2}  & 5.43 &     0.97 &                   $-$0.58 &    0.03 &   4 \\
\ion{Fe}{1}  & 7.50 &     3.62 &  $-$3.88\tablenotemark{a} &    0.01 &  98 \\
\ion{Fe}{2}  & 7.50 &     3.63 &  $-$3.87\tablenotemark{a} &    0.01 &  53 \\
\ion{Ni}{1}  & 6.22 &     2.32 &                   $-$0.02 &    0.02 &  19 \\
\ion{Ni}{2}  & 6.22 &     2.21 &                   $-$0.13 &    0.03 &   8 \\
\ion{Ge}{1}  & 3.65 & $<-$0.80 &                  $<-$0.57 & \nodata &   3 \\
\ion{Zr}{2}  & 2.58 & $<-$1.11 &                  $<+$0.18 & \nodata &   7 \\
\ion{Nb}{2}  & 1.46 & $<-$0.77 &                  $<+$1.65 & \nodata &   2 \\
\ion{Mo}{2}  & 1.88 & $<-$0.29 &                  $<+$1.71 & \nodata &   1 \\
\ion{Cd}{1}  & 1.71 & $<-$1.79 &                  $<+$0.38 & \nodata &   1 \\
\ion{Te}{1}  & 2.18 & $<-$0.05 &                  $<+$1.65 & \nodata &   1 \\
\ion{Ce}{2}  & 1.58 & $<-$0.51 &                  $<+$1.79 & \nodata &   1 \\
\ion{Nd}{2}  & 1.42 & $<-$0.01 &                  $<+$2.45 & \nodata &   1 \\
\ion{Eu}{2}  & 0.52 & $<-$1.46 &                  $<+$1.90 & \nodata &   1 \\
\ion{Gd}{2}  & 1.07 & $<-$1.21 &                  $<+$1.60 & \nodata &   1 \\
\ion{Yb}{2}  & 0.84 & $<-$1.75 &                  $<+$1.29 & \nodata &   1 \\
\ion{Hf}{2}  & 0.85 & $<-$1.31 &                  $<+$0.00 & \nodata &   1 \\
\ion{Lu}{2}  & 0.10 & $<-$2.16 &                  $<+$1.62 & \nodata &   1 \\
\ion{Hf}{2}  & 0.85 & $<-$1.31 &                  $<+$1.72 & \nodata &   3 \\
\ion{Os}{1}  & 1.40 & $<-$0.30 &                  $<+$2.18 & \nodata &   1 \\
\ion{Os}{2}  & 1.40 & $<-$0.90 &                  $<+$1.58 & \nodata &   1 \\
\ion{Pt}{1}  & 1.62 & $<-$1.40 &                  $<+$0.86 & \nodata &   2 \\
\ion{Pb}{1}  & 1.75 & $<-$0.23 &                  $<+$1.90 & \nodata &   1
\enddata
\tablenotetext{a}{$\mbox{[\ion{Fe}{1}/H]}$ and $\mbox{[\ion{Fe}{2}/H]}$ values}
\end{deluxetable}

\begin{deluxetable}{lcccc}
\tablewidth{0pc}
\tablecaption{Systematic Abundance Uncertainties \label{sys}}
\tablehead{
\colhead{Species}&
\colhead{$\Delta$\teff}&
\colhead{$\Delta$\logg}&
\colhead{$\Delta v_\mathrm{micro}$}&
\colhead{$\sigma_{\rm tot}$}\\
\colhead{}&
\colhead{$+$150\,K}&
\colhead{$+$0.5 dex}&
\colhead{$+$0.3 km/s}&
\colhead{}}
\startdata
\ion{Sc}{2} & $-$0.10 & $-$0.17 & $+$0.02 & 0.20 \\ 
\ion{Ti}{2} & $-$0.09 & $-$0.17 & $+$0.04 & 0.20 \\ 
\ion{Cr}{2} & $-$0.07 & $-$0.16 & $+$0.11 & 0.21 \\ 
\ion{Mn}{2} & $-$0.12 & $-$0.06 & $+$0.13 & 0.19 \\ 
\ion{Fe}{1} & $-$0.18 & $+$0.02 & $+$0.06 & 0.19 \\ 
\ion{Fe}{2} & $-$0.05 & $-$0.15 & $+$0.05 & 0.17 \\ 
\ion{Ni}{1} & $-$0.20 & $+$0.02 & $+$0.09 & 0.22 \\ 
\ion{Ni}{2} & $-$0.06 & $-$0.17 & $+$0.14 & 0.23
\enddata
\end{deluxetable}

\end{document}